\documentclass[12pt,a4paper]{article}
\usepackage{color, graphicx}

\def\lvec#1{\setbox0=\hbox{$#1$}
    \setbox1=\hbox{$\scriptstyle\leftarrow$}
    #1\kern-\wd0\smash{
    \raise\ht0\hbox{$\raise1pt\hbox{$\scriptstyle\leftarrow$}$}}
    \kern-\wd1\kern\wd0}
\def\rvec#1{\setbox0=\hbox{$#1$}
    \setbox1=\hbox{$\scriptstyle\rightarrow$}
    #1\kern-\wd0\smash{
    \raise\ht0\hbox{$\raise1pt\hbox{$\scriptstyle\rightarrow$}$}}
    \kern-\wd1\kern\wd0}

\def\diracstar#1#2{
    \setbox0=\hbox{$\gamma$}\setbox1=\hbox{$\gamma_{#1}$}
    \gamma_{#1}\kern-\wd1\kern\wd0
    \smash{\raise4.5pt\hbox{$\scriptstyle#2$}}}

\newcommand{\beq}{\begin{equation}}
\newcommand{\eeq}{\end{equation}}
\newcommand{\beqn}{\begin{eqnarray}}
\newcommand{\eeqn}{\end{eqnarray}}
\newcommand{\nn}{\nonumber}

\begin{document} 

\begin{titlepage}

\title{\bf The string-junction picture of multiquark states: an update}

\author{G.C.~Rossi$^{a)}$\, and\, G.~Veneziano$^{b)}$}
\date{}

\maketitle

\begin{center}
$^{a)}${\small Dipartimento di Fisica, Universit\`a di Roma {\it Tor Vergata} \\
INFN, Sezione di Roma 2 \\ Via della Ricerca Scientifica, 00133 Roma, Italy \\ 
Centro Fermi - Museo Storico della Fisica\\ Piazza del Viminale 1 - 00184 Roma, Italy} \\
\vspace{.1cm}
$^{b)}${\small  Coll\`ege de France, 11 place M. Berthelot, 75005 Paris, France \\
Theory Division, CERN, CH-1211 Geneva 23, Switzerland \\ Dipartimento di Fisica, Universit\`a di Roma La Sapienza, 00185 Rome, Italy}
\end{center}

\begin{abstract}
We recall and update, both theoretically and phenomenologically, our (nearly) forty-years-old proposal of a string-junction as a necessary complement to the conventional classification of  hadrons based just on their quark-antiquark constituents. In that proposal single (though in general metastable) hadronic states are associated with ``irreducible" gauge-invariant operators consisting of Wilson lines (visualized as strings of color flux tubes)  that  may either end on a quark or an antiquark, or annihilate in triplets at a junction $J$ or an anti-junction $\bar{J}$. For the junction-free sector (ordinary $q\, \bar{q}$ mesons and glueballs) the picture is supported by large-$N$ (number of colors) considerations as well as by a lattice strong-coupling expansion. Both imply the famous OZI rule suppressing quark-antiquark annihilation diagrams. For hadrons with $J$ and/or $\bar{J}$ constituents the same expansions support our proposal, including its generalization of the OZI rule to the suppression of $J-\bar{J}$ annihilation diagrams. Such a rule implies that hadrons with junctions are ``mesophobic" and thus unusually narrow if they are below threshold for decaying into as many baryons as their total number of junctions (two for a tetraquark, three for a pentaquark). Experimental support for our claim, based on the observation that narrow multiquark states typically lie below (well above) the relevant baryonic (mesonic) thresholds, will be presented. 

\end{abstract}

\vspace{-23.cm}
\flushright{CERN-TH-2016-053}

\end{titlepage}

\newpage

\section{Introduction}

The prediction of multiquark states predates the QCD era. Soon after the Dolen--Horn--Schmidt (DHS) proposal~\cite{DHS} of a duality between Regge poles in the $t$ channel and resonances in the $s$ and $u$ channels, Rosner~\cite{ROSNER} pointed out that a straightforward application of that concept to baryon-antibaryon scattering implied the existence of tetraquark (exotic) states dual to the exchange of ordinary quark-antiquark mesons. Early claims of their actual existence, however, turned out to be either unfounded or inconclusive. Convincing evidence in favor of their existence is relatively recent: what seems to have put the whole field on solid grounds is the discovery of multiquark states containing heavy ($c$ or $b$) quarks. This has brought renewed interest in the subject and, in particular, on the question of how to interpret this new class of hadrons within QCD.

For processes involving just mesons the connection between DHS-duality related arguments and QCD was put on solid grounds through the introduction of large-$N$ expansions of generalizations of QCD  to an arbitrary number $N$ of colors and $N_f$ of flavors. In particular, 't Hooft's expansion~\cite{'tHooft} makes a precise connection between the leading QCD diagrams and the duality diagrams of DHS. Those considerations can be extended~\cite{GVTE} to the glueball sector of QCD confirming, in particular, the duality connection  between glueball (Pomeron) exchange and a non-resonant two-meson background~\cite{Harari,Freund}. Apparently, these two-particle states had nothing to do with Rosner's original tetraquarks, but how does QCD make the distinction?

Some 40 years ago~\cite{Rossi:1977cy} (see also the review paper~\cite{Montanet:1980te} covering as well the experimental situation at the time) we did try to reinterpret Rosner's original observation within QCD. Rosner's states were dubbed ``baryonium" (for hidden-baryon-number states) for reasons that will be clarified below. The starting point of the analysis of ref.~\cite{Rossi:1977cy} was the association of {\it single} (stable or metastable) hadrons with {\it irreducible} gauge invariant operators in QCD~\footnote{(Irr)reducibility will thus distinguish tetraquarks from the above two-meson states of refs.~\cite{Harari,Freund}.}. These were taken to be in one-to-one correspondence with {\it connected} graphs made of lines which could either end at a quark ($q$) or an antiquark ($\bar{q}$), or join in triplets at a junction ($J$) or an antijunction ($\bar{J}$). The lines were nothing but the Wilson lines (path-ordered exponentials of the gauge connection) needed for gauge invariance. Examples will be given in the corresponding appropriate sections. 

The main novelty of our proposal was the necessity of introducing $J$ and $\bar{J}$ as new essential constituents for a complete and unambiguous classification of single hadron states as well as of processes involving baryons. Of course, the possibility of having gauge-invariant operators with junctions is related to the fact that the QCD gauge group is $SU(3)$, rather than $U(3) = SU(3) \otimes U(1)$.
A couple of years later Witten~\cite{Witten_Baryon} went further by discussing the systematics of the large-$N$ expansion for baryons reaching the interesting conclusion that baryons behave like solitons since their mass is proportional to $N \sim 1/g^2$. He also discussed several other features of baryons at large-$N$ (with and without use of the string-junction picture) that we shall refer to where appropriate in the rest of the paper.

We would like to stress immediately that our approach to multiquark states differs in a substantial way from other schemes independently proposed at about the same time~\cite{VO}-\cite{STR}, as well as from later constructions like~\cite{LIP}-\cite{JW} or~\cite{MPPR1}-\cite{Maiani:2015vwa} (see also the review~\cite{Esposito:2014rxa} and references therein). We concentrate on the big family of multiquark states endowed with junctions of which the ordinary baryons (states with one junction) represent the simplest sector. 

Members of this  family strongly interact with each other: they should be better called baryonia. Likewise, hadrons without junctions (ordinary mesons as well as other possible multiquark states) strongly couple among themselves, while the mutual interactions between the two sectors (more generally between sectors with a different total number of junctions) are suppressed. Thus, in our picture, any large-$N$ extrapolation of multiquark states has to follow the ``baryonic route" of keeping the number of junctions fixed.

In this paper  we would like to (recall and) update, both theoretically and phenomenologically, the proposals of~\cite{Rossi:1977cy} and~\cite{Witten_Baryon} according to the following outline. In Section~\ref{sec:JFS} we briefly review the standard description of $q\, \bar{q}$ mesons and glueballs and of their properties both in the large-$N$ limit~\cite{'tHooft} and in the strong-coupling expansion~\cite{strongcoupling} of lattice QCD (LQCD). The latter is argued to become, at large-$N$, a large $\lambda\equiv g^2 N$ ('t Hooft's coupling) expansion valid even at small $g^2$. Particular attention is paid to the famous OZI rule~\cite{OKU,ZWE,IIZ} that suppresses mixing of these two classes of states and is responsible for the narrow width of several quarkonium states. 
In Section~\ref{sec:BATE} we turn our attention to baryons and to baryon-(anti)baryon scattering amplitudes arguing that, in QCD, single tetraquark states should exist although they can mix via $J-\bar{J}$ annihilation with ordinary (single or multi)-meson states. After a short reminder of Witten's large-$N$ expansion for baryons~\cite{Witten_Baryon} we turn to the strong-coupling and large-$\lambda$ expansions. We will argue that all these approaches neatly show the emergence of a junction in a ``baryonic Wilson loop" (simulating a baryon propagator for large quark masses) and imply the distinction between scattering and annihilation channels in baryon-antibaryon collisions. 
In Section~\ref{sec:JOZI} we recall the so-called junction-OZI (JOZI) rule proposed in~\cite{Rossi:1977cy}, by which, for instance, tetraquarks prefer to decay into baryon-antibaryon channels whenever this is allowed by phase space, and should otherwise be unusually narrow. After offering some theoretical justification for the JOZI rule from the large-$N$ and large-$\lambda$ expansions we briefly review experimental evidence for its validity both for tetraquarks and pentaquarks. In Section~\ref{sec:CONC} we summarize our conclusions. Some technical details concerning Witten's large-$N$  expansion for baryons are relegated to an appendix.

\section{The junction-free sector}
\label{sec:JFS} 

\subsection {Ordinary $q\, \bar{q}$ mesons and glueballs}

In our approach ordinary $q\, \bar{q}$  mesons are associated with the irreducible (single trace) gauge-invariant operator 
\beqn
&&{\cal M}({\cal C}_t)  =  \frac{1}{\sqrt{N}}\bar q_i(\vec r,t) U[{\cal C}_t]_j^i q^j(\vec s,t) 
\label{MESON}
\eeqn
where the Wilson-line operator $U$ is defined by
\beq
U[{\cal C}_t]_j^i ={\cal P} \exp\Big{[}ig\int_{\vec r}^{\vec s}d\vec x\vec A(\vec x,t) \Big{]}_j^i
\label{WLINE}
\eeq
with ${\cal C}_t$ a line joining the point ${\vec r},t$ with ${\vec s},t$. More generally, we may consider as an interpolating operator one in which the path joins two arbitrary space-time points. But the above class of paths is sufficient for our purposes  and is easier to consider in the strong-coupling limit. 

Ordinary $q\, \bar{q}$  mesons will appear as intermediate states in the gauge invariant correlator (after subtracting a disconnected contribution in the flavor-singlet channel)
 \beq
G_{\cal M}({\cal C}_{t'},{\cal C}_t)= \langle {\cal M}({\cal C}_{t'})  {\cal M}^\dagger({\cal C}_t) \rangle\, .
\label{MP1}
\eeq
Similarly, glueballs are associated with the irreducible (single trace) gauge-invariant operator 
\beq
{\cal G}({\cal C}_t) =  {\rm Tr}{\cal P} \exp\Big{[}ig\oint_{\cal C}d\vec x\vec A(\vec x,t) \Big{]}
\label{WGLINE}
\eeq
and will appear as intermediate states in the (connected part of the) correlator 
\beq
G_{\cal G}({\cal C}_{t'},{\cal C}_t)= \langle {\cal G}({\cal C}_{t'})  {\cal G}^\dagger({\cal C}_t) \rangle\, ,
\label{MG1}
\eeq
where ${\cal C}_t$ is now a closed spatial loop at time $t$. 

However, there is no strict selection rule preventing single-trace operators from mixing with multi-trace operators and for flavor-singlet $q\, \bar{q}$ mesons to mix with the glueballs. In full QCD the above correlators will have singularities in correspondence with the exact spectrum of QCD (including widths, branch points etc.). 

Fortunately, the situation simplifies enormously by considering two limits: 't Hooft's large-$N$ limit with fixed $ \lambda \equiv g^2 N$ and, on the lattice, the strong-coupling limit, $g^2\to \infty$ with fixed $N$ as well as a large-$\lambda$ limit in which $N$ is large but $g^2$ can be small.

\subsection{Two simplifying limits and the OZI rule}
\label{sec:OZIR}

\subsubsection {The large-$N$ limit}

The simplifications occurring in 't Hooft's large-$N$  limit~\cite{'tHooft} (with fixed $g^2N$ and $N_f$, see also~\cite{GVTE}~\cite{Witten_Baryon}) are well known. 
\begin{itemize}
\item Color irreducible operators do not mix with reducible ones. Consequently the states we have introduced above are stable (zero-width) hadrons in the large-$N$ limit.
\item This is confirmed by the fact that the coupling among $n$ -- $q\, \bar{q}$ mesons goes to zero like $N^{1-n/2}$. This result can be easily generalized \cite{GVTE} to multi-glueball couplings (that scale like $N^{2-n}$) and to mixed ones (that scale like $N^{1-n_{q \bar{q}}/2-n_{gl}}$).
\item As a consequence of the previous properties there is no mixing between $q\, \bar{q}$ mesons and glueballs. They represent two decoupled sectors of stable mesons. Also, there is no mixing among quarkonia of different flavor, a property that can induce, in principle, large isospin mixing in some multiquark states~\cite{Rossi:1977dp,Rossi:2004yr}.
\item The absence of mixing with glueballs reflects the validity of the OZI rule at large-$N$: $q\, \bar{q}$ pairs do not annihilate in the 't Hooft limit!
As an example  of the implications of the OZI rule an $s \bar{s}$ meson prefers to decay into a strange pair, a $c \bar{c}$ meson prefers to decay into a charmed pair, etc. If such decays are kinematically forbidden the state is unusually narrow.
\item The validity of the OZI rule at leading order in $N$ also implies the absence, at the same order, of the $U(1)_A$ anomaly. Thus, at this order, the flavor-singlet pseudo scalar is a true (pseudo)-Nambu--Goldstone boson. At next to leading order one instead derives a successful formula~\cite{Witten:1979vv}~\cite{Veneziano:1979ec} for the mass of the  $\eta'$ meson.
\item The leading (planar) diagrams for the scattering of  $q\, \bar{q}$ mesons will exhibit (under mild assumptions about their large-$s$, fixed-$t$ Regge limit) the usual planar duality between $s$, $t$ and $u$ channels and their generalizations to higher-point functions. At next to leading order, there are also non-planar diagrams contributing to $q\, \bar{q}$  meson scattering. They will exhibit duality~\cite{Harari,Freund} between glueball states in the $t$ channel and states consisting of two $q\, \bar{q}$ mesons (thus corresponding to reducible color singlet operators) in the $s$ and $u$ channels (see also the discussion at the end of Section~\ref{sec:SCLL}).
\end{itemize}

\subsubsection {The strong coupling and large-$\lambda$ limits on the lattice}
\label{sec:SCLL}

An interesting alternative to the large-$N$ expansion can be defined and nicely implemented on the lattice. This is the so-called strong coupling expansion~\cite{strongcoupling,Drouffe:1983fv,OBZ}: it bears some interesting analogies with the large-$N$ expansion and, as we shall discuss, can be combined with it in a large-$\lambda$ expansion. Its advantage is that, in many cases, the leading term can be explicitly computed analytically and is exactly gauge invariant.

On the negative side we know that such an expansion can only give some qualitative information about the true continuum theory which, because of asymptotic freedom, corresponds to a vanishing bare ('t Hooft) coupling (to be identified with the coupling at the lattice cutoff scale). Even getting correct qualitative information is not guaranteed. A well-known example is a $U(1)$ gauge theory (e.g.\ QED~\cite{JNZ1,DKK}) which confines at strong coupling while it is in the Coulomb phase in the continuum, because of the existence of a first-order phase transition at a finite value of the coupling constant.

In the strong coupling limit it is more convenient to talk about Wilson loops. They naturally emerge as soon as quark propagators are replaced (in the large mass limit) by a  Wilson line in the time direction (i.e.\ at a fixed spatial position). 

\begin{figure}[htbp]
\centerline{\includegraphics[scale=0.8,angle=0]{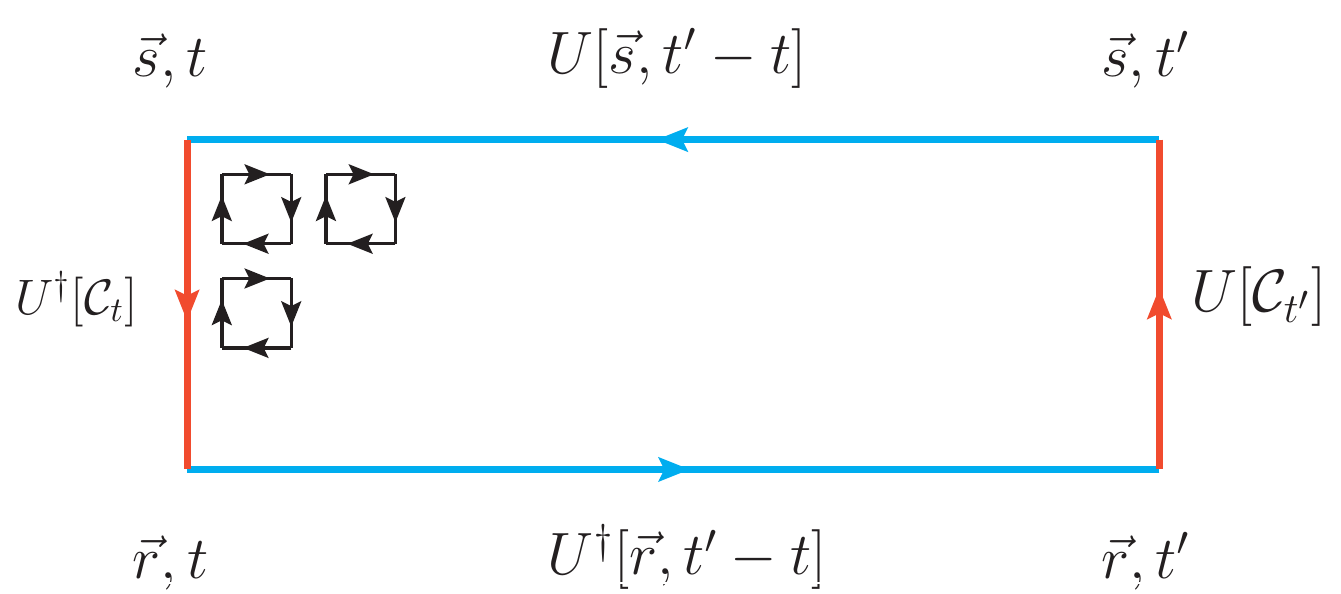}}
\caption{\small{Meson propagator in the strong coupling limit. The Wilson line connecting $q$ with $\bar q$ is in red. The quark propagators are in light blue.}}
\label{fig:fig1}
\end{figure}

1) As a first example consider the connected meson propagator~(\ref{MP1}) on the lattice. Contracting the quark fields one finds  
\beqn
\hspace{-.8cm}&&G_{\cal M}({\cal C}_{t'},{\cal C}_t) = \nn\\
\hspace{-.8cm}&&=\frac{1}{N} \frac{\int \prod_{i} \!dU_i {\rm Tr}\Big{(}  U^\dagger[{\cal C}_t]S_F(\vec r,t;\vec r, t') U[{\cal C}_{t'} ] S_F(\vec s, t';\vec s,t) \Big{)}e^{-\frac{1}{g^2}S_{LYM}(U)}}{\int \prod_{i} dU_i  e^{-\frac{1}{g^2}S_{LYM}(U)}}\, .
\label{MP2}
\eeqn
Since in the limit of a very massive quark (static limit) we can replace the quark propagator with the product of the links in the time direction from $t$ to $t'$, we end up with the correlator 
\beqn
\hspace{-.4cm}&&G_{\cal M}({\cal C}_{t'},{\cal C}_t)\Big{|}_{large\, mass}=\nn\\
\hspace{-.4cm}&&=\frac{1}{N}\frac{\int \prod_{i} dU_i\,{\rm Tr}\Big{(}  U^\dagger[{\cal C}_{t}]U^\dagger[\vec r,t-t'] U[{\cal C}_{t'} ] U[\vec s,t'-t] \Big{)}e^{-\frac{1}{g^2}S_{LYM}(U)}}{\int \prod_{i} dU_i  e^{-\frac{1}{g^2}S_{LYM}(U)}}=\nn\\
\hspace{-.4cm}&&=\frac{1}{N}\langle {\rm Tr}\Big{(}  U^\dagger[{\cal C}_{t}]U^\dagger[\vec r,t-t'] U[{\cal C}_{t'} ] U[\vec s,t'-t] \Big{)} \rangle \equiv W_{\cal M} \, ,
\label{MP3}
\eeqn
where $S_{LYM}(U)$ is the lattice pure gauge action and
\beq
U[\vec s,t'-t] =\prod_{\tau \in[t,t']} U[\vec s,\tau]\, .
\label{LINT}
\eeq
Looking at fig.~\ref{fig:fig1}, one recognizes in $G_M({\cal C}_{t'},{\cal C}_t)$ the expectation value of a Wilson loop with sides $|\vec s -\vec r|\times |t'-t|$.

For a generic $N$  the strong coupling expansion is defined as the one in which $N$ is kept fixed while $\beta \equiv 2N/g^2 \rightarrow 0$. Carrying out the actual calculation for the rectangular Wilson loop of fig.~\ref{fig:fig1} and using the standard group integration rules~\cite{Creutz:1978ub}, one finds from~(\ref{MP3})
\beqn
&&\lim_{Strong \,\, Coupling} W_{\cal M}  \propto \exp[-A/a^2 \log(g^2N)]\label{WMES}\\
&&A=|\vec s -\vec r|\times |t'-t| \, .\label{AREAM}
\eeqn
The result~(\ref{WMES}) is obtained by bringing down from the action the minimal number of plaquettes allowing to have a non-vanishing group integral. This amounts to tiling with plaquettes the rectangular Wilson loop of fig.~\ref{fig:fig1}.

We thus see that in the case of a $q\, \bar{q}$ meson the area is just $|\vec s -\vec r|\times |t'-t|\equiv L T$. Therefore, interpreting the coefficient of $T$ as the energy and the energy as the tension, $\kappa$, times the distance $L$, we find 
\beq
\kappa=\frac{1}{a^2}\log g^2N \, .
\label{WILDEF}
\eeq
An important observation here is that, actually, the leading strong coupling term does not depend on $g^2$ and $N$ separately but only on their combination $\lambda = g^2N$ where, in the naive continuum limit, $\lambda$ is nothing but the 't Hooft coupling. This conclusion can be extended to the subleading terms~\cite{Zpc} (after removing disconnected diagrams) and the whole strong coupling expansion can be rearranged in the 't Hooft form~\cite{OBZ}
\beq
W_{\cal M} =\sum_{h=0} W_h(\lambda)\frac{1}{(N^2)^h}\, ,
\label{LAMN}
\eeq
where the sum is over the number of handles, $h$, of the diagram. The expansion~(\ref{LAMN}) tells us that the strong coupling expansion is actually a large-$\lambda$ ('t-Hooft-coupling) expansion, i.e. the corrections to the leading term scale like powers of $\lambda^{-1} = \frac{1}{g^2 N}$ and $1/N^2$, and {\it not} of $1/g^2$. Therefore such an expansion is also valid at small $g^2$ provided $\lambda$ and thus $N\gg1$~\footnote{This limit is  like the much used large AdS-radius limit of the AdS/CFT  correspondence~\cite{Maldacena:1997re,Witten:1998qj}. In ref.~\cite{Andreev:2015riv} the possibility of establishing a bridge between the stringy description of QCD resulting from the AdS/CFT correspondence and the strong coupling limit of lattice QCD in the study of the potential among the quarks of a triply heavy baryon is explored.}. This observation will be relevant for the extension of our considerations to baryons and multiquark states.

2) As a next step we can similarly compute the glueball-glueball correlator in the strong coupling limit. Starting from the definition~(\ref{WGLINE}) with the integral extended over the closed curve, ${\cal C}$, that for simplicity we have assumed to lie in a plane at time $t$, one finds that in the $\beta\to 0$, fixed $N$-limit the dominant topology is that of a cylinder (actually a parallelepiped) whose bases are the two parallel closed curves ${\cal C}_t$ and ${\cal C}_{t'}$,  and the height is $|t-t'|$. One thus gets
\beqn
\hspace{-1.cm}&& \lim_{Strong \,\, Coupling}  G_{\cal G}({\cal C}_{t'},{\cal C}_t)\equiv \lim_{Strong \,\, Coupling} W_{\cal G} 
\propto \exp[-A/a^2 \log(g^2N)]\, ,\label{WGLUEB} \\
\hspace{-1.cm}&& \nn\\
\hspace{-1.cm}&& \qquad A=2(L+d) |t-t'| \sim 2LT\, ,\qquad d\ll L\, ,\label{WGLUEBA}
\eeqn
where the condition $d\ll L$ means that we are looking at states with large angular momentum. We thus find that the effective ``glueball string tension'', defined as the inverse of the Regge slope, is twice as large as the mesonic one, in full agreement with the string picture (naturally the energy per unit string length is always the same and does not depend on whether one is dealing with mesonic or gluonic states). 

\begin{figure}[htbp]
\centerline{\includegraphics[scale=0.8,angle=0]{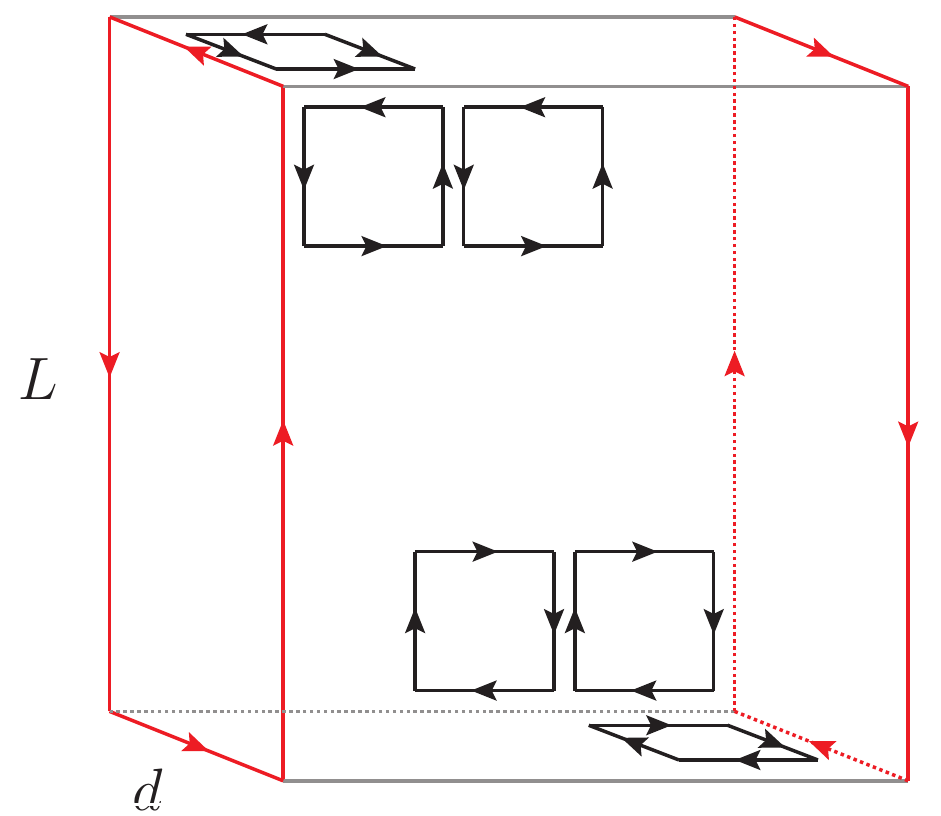}}
\caption{\small{Glueball propagator in the strong coupling limit. The curves ${\cal C}_t$ and ${\cal C}_{t'}$ are in red. The gray horizontal lines are only drawn to guide the eye.}}
\label{fig:fig2}
\end{figure}

We now argue that, like in the large-$N$ expansion, also the strong coupling expansion implies the validity of the OZI rule. We will concentrate on an important consequence of it, namely the absence of mixing between two quarkonium states carrying different hidden flavor (which we already mentioned as a property of the large-$N$ expansion). 

\begin{figure}[htbp]
\centerline{\includegraphics[scale=0.8,angle=0]{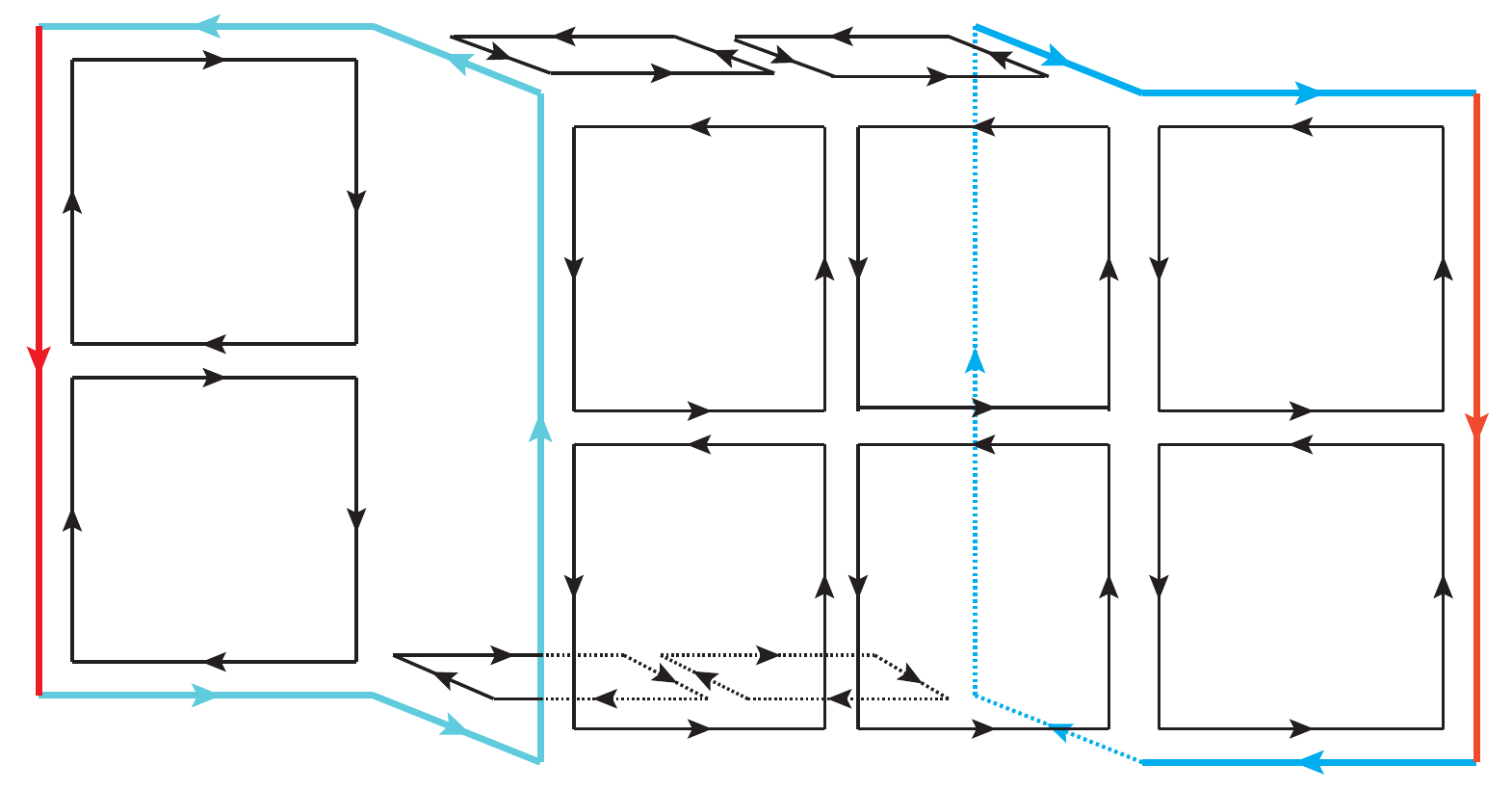}}
\caption{\small{Meson-to-meson mixing. As in fig.~\ref{fig:fig1} the Wilson lines connecting $q$ with $\bar q$ is in red and the quark propagators in blight blue.}}
\label{fig:fig3}
\end{figure}

Let us compare the (OZI-conserving) two-point function of fig.~\ref{fig:fig1} with the OZI-violating one depicted in fig.~\ref{fig:fig3}. While the former behaves as 
$\exp[-\kappa A]$ (see eq.~(\ref{AREAM}) ), where $A$ is the area of the rectangle, in the latter case we have to insert, somewhere in the strong coupling diagram, a complicated non-planar feature of the kind shown in fig.~\ref{fig:fig3}. It is then clear that, for a fixed $L$ and $T$, this feature can only increase the minimal area needed to properly tile the diagram. Basically, during a certain time-interval, we will be dealing with a glueball propagator and thus pay the price of a larger tension. Consequently the OZI-violating contribution will contain extra inverse powers of $\lambda$.

\begin{figure}[htbp]
\centerline{\includegraphics[scale=0.6,angle=0]{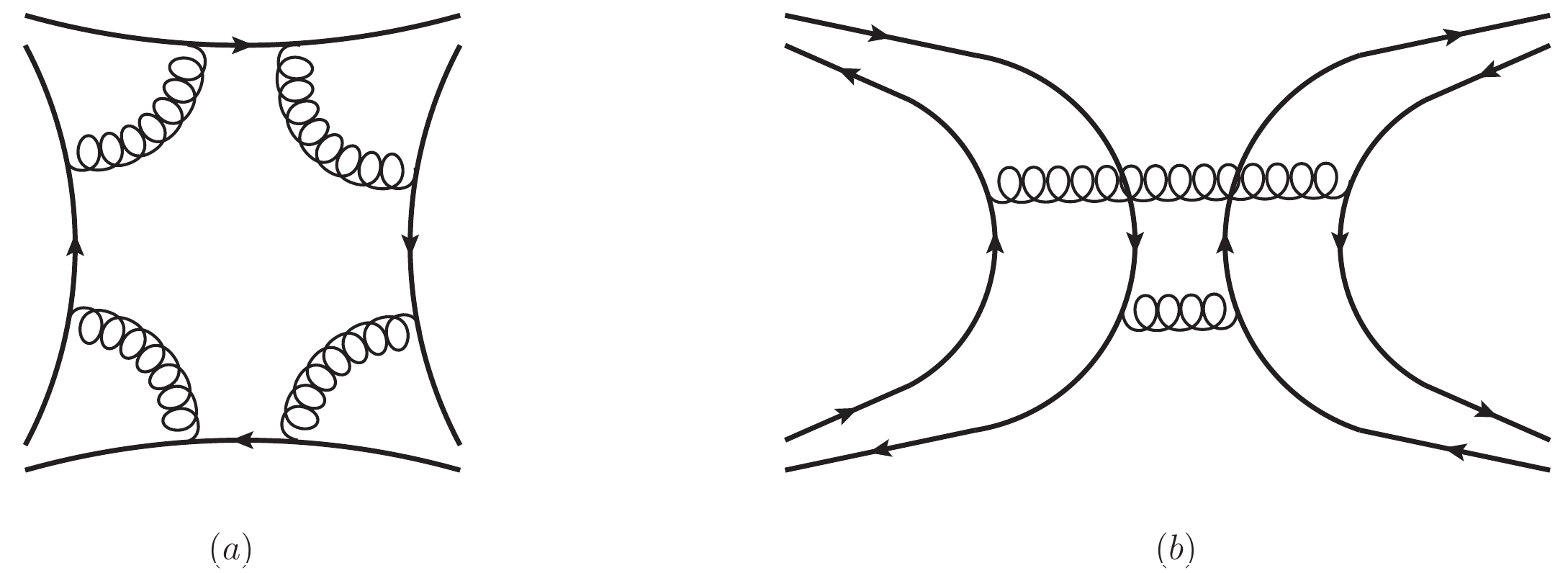}}
\caption{\small{Two contributions to meson-meson scattering in the large-$N$ limit. Panel (a) is the leading OZI-preserving  term; panel (b) is a non-planar OZI-violating subleading correction. But at sufficiently high energy in the crossed channel, (b) dominates over (a) because of the higher intercept of the flavor singlet Regge pole.
}}
\label{fig:MM2}
\end{figure}

Before moving on to baryons we wish to make a point that will be relevant later. 
Consider the two contributions to meson-meson scattering depicted in fig.~\ref{fig:MM2}. The non-planar diagram in the panel (b) is obviously subleading at large $N$. It is also subleading at high energy for the channel in which the OZI rule is violated. This is because the leading Regge singularity in its $t$ channel is a two-Reggeon cut, which is certainly lower than the single $q\, \bar{q}$ Regge pole exchanged in fig.~\ref{fig:MM2}a. Let us now look at the diagram of fig.~\ref{fig:MM2}b from the crossed channel viewpoint, in which all four quark lines go through. Since the diagram is always the same it is still down, at large $N$, with respect to the planar diagram of fig.~\ref{fig:MM2}a. However, the diagram of  fig.~\ref{fig:MM2}b is now {\it dominant} at high energy, since it allows for the exchange of the leading (vacuum) Pomeron trajectory in its own $t$ channel. This shows that there is sometimes competition between large $N$ and high-energy approximations. We shall see something analogous to this when discussing the JOZI rule in Section~\ref{sec:JOZI}.

\section{Baryons, junctions and tetraquarks}
\label{sec:BATE}

\subsection {Single baryon states}

\begin{figure}[htbp]
\centerline{\includegraphics[scale=0.8,angle=0]{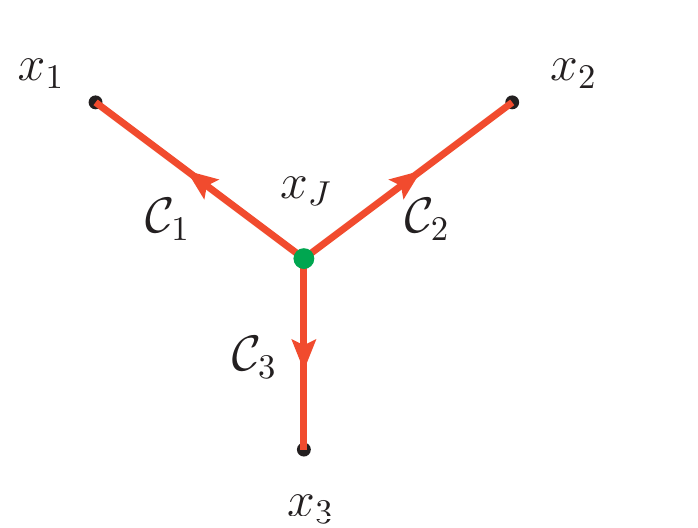}}
\caption{\small{The Y-shaped form of the baryon for $N=3$.}}
\label{fig:fig4}
\end{figure}

In SU($N$) QCD the (normalized) irreducible gauge invariant operator of a baryon takes the Y-shaped (for $N=3$) form (see fig.~\ref{fig:fig4})
\beqn
\hspace{-.8cm}&&B({\cal C}_1,{\cal C}_2,\ldots,{\cal C}_N)=\nn\\
\hspace{-.8cm}&&=\frac{1}{\sqrt{N!}} \epsilon_{i_1i_2\ldots i_N} U[{\cal C}_1]^{i_1}_{j_1} q(x_1)^{j_1}\, U[{\cal C}_2]^{i_2}_{j_2}q(x_2)^{j_2}\ldots U[{\cal C}_N]^{i_N}_{j_N}q(x_N)^{j_N} \, ,
\label{BARY}
\eeqn
where
\beq
U[{\cal C}_k]^{i_k}_{j_k}={\cal P}\exp \Big{[}ig\int_{{\cal  C}(x_J,x_k)} dy^\mu A_\mu(y) \Big{]}^{i_k}_{j_k}\, , \quad k=1,2,\ldots,N
\label{UDEF}
\eeq
and ${\cal  C}(x_J,x_k)$ is a curve joining the point $x_J$ to $x_k$. As in the mesonic case, we have taken for simplicity very special space-time locations for the $q$ fields. The description of baryons as a triplet of flux tubes joining at a point dates back to the work of ref.~\cite{Artru:1974zn}, where the word ``junction'' was first introduced (see also~\cite{Isgur:1984bm}). 

Single baryon intermediate states appear in the correlator
\beqn
\hspace{-.8cm}&&G_B(\{\vec r_k, k=1,2,\ldots, N\}, \vec r_J; t'-t) =\nn\\
\hspace{-.8cm}&&= \langle B({\cal C}_1,{\cal C}_2,\ldots,{\cal C}_N) B^\dagger({\cal C}'_1,{\cal C}'_2,\ldots,{\cal C}'_N)\rangle \, .
\label{BARPROP}
\eeqn
We will now discuss how the treatment of this correlator simplifies in the large-$N$ and strong coupling limit of LQCD, starting with the latter.
 
\subsection{Strong coupling, large-$\lambda$ considerations}

\begin{figure}[htbp]
\centerline{\includegraphics[scale=0.8,angle=0]{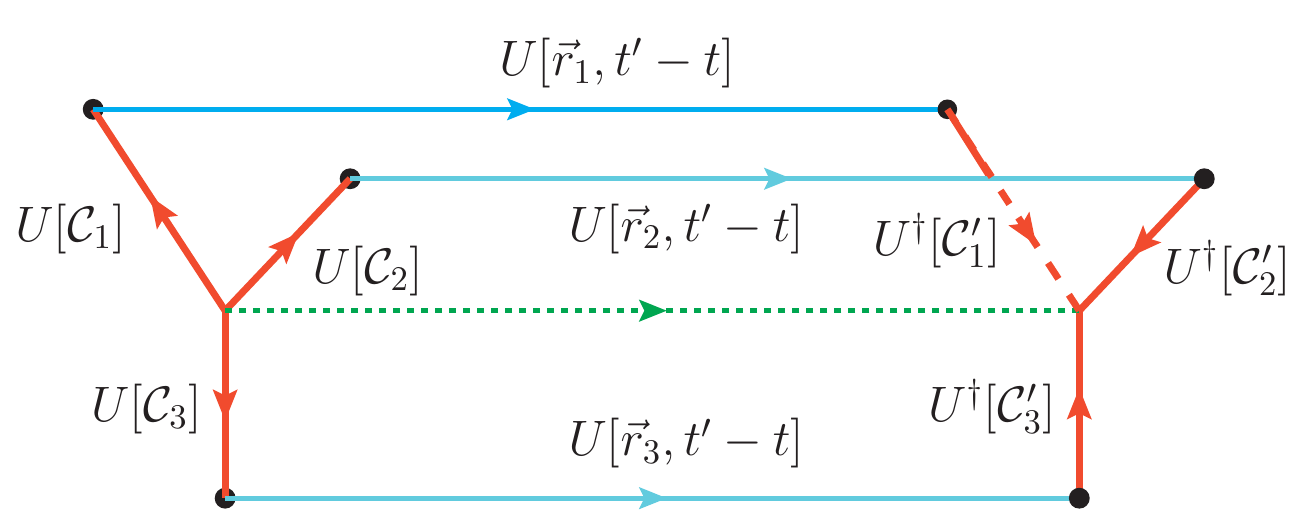}}
\caption{\small{The $N=3$ baryonic Wilson loop. The green dotted line does not explicitly appear in the correlator~(\ref{PROLM}) but it will come out from the calculation outlined below.}}
\label{fig:fig5}
\end{figure}

Putting $x_k=(\vec r_k,t),  k=1,2,\ldots ,N ; x_J=(\vec r_J,t)$ and similarly $x'_k=(\vec r_k,t')$, $k=1,2,\ldots ,N ; x'_J=(\vec r_J,t')$, we want to evaluate $G_B$ in the strong coupling limit. Following the strategy outlined in sect.~\ref{sec:SCLL} in the case of the meson propagator leads to a new kind of Wilson loop, the baryonic Wilson loop depicted in fig.~\ref{fig:fig5}, characterized by the presence of the Levi-Civita symbol. It reads~\cite{Brambilla:1993zw,Kalashnikova:1996px} 
\beqn
\hspace{-.8cm}&&G_B(\{\vec r_k, k=1,2,\ldots, N\}, \vec r_J; t'-t)\Big{|}_{large\,mass} =\frac{1}{N!} \epsilon_{i_1\ldots i_N} \epsilon^{i'_1\ldots i'_N}\cdot \label{PROLM}\\
\hspace{-.8cm}&&\cdot \langle \,U[{\cal C}_1]^{i_1}_{j_1} U[\vec r_1,t-t']^{j_1}_{j'_1} U^\dagger[{\cal C}'_1]^{j'_1}_{i'_1} \ldots U[{\cal C}_N]^{i_N}_{j_N} U[\vec r_N,t-t']^{j_N}_{j'_N} U^\dagger[{\cal C}'_N]^{j'_N}_{i'_N}\,\rangle \equiv W_J \, . \nn
\eeqn

\begin{figure}[htbp]
\centerline{\includegraphics[scale=0.8,angle=0]{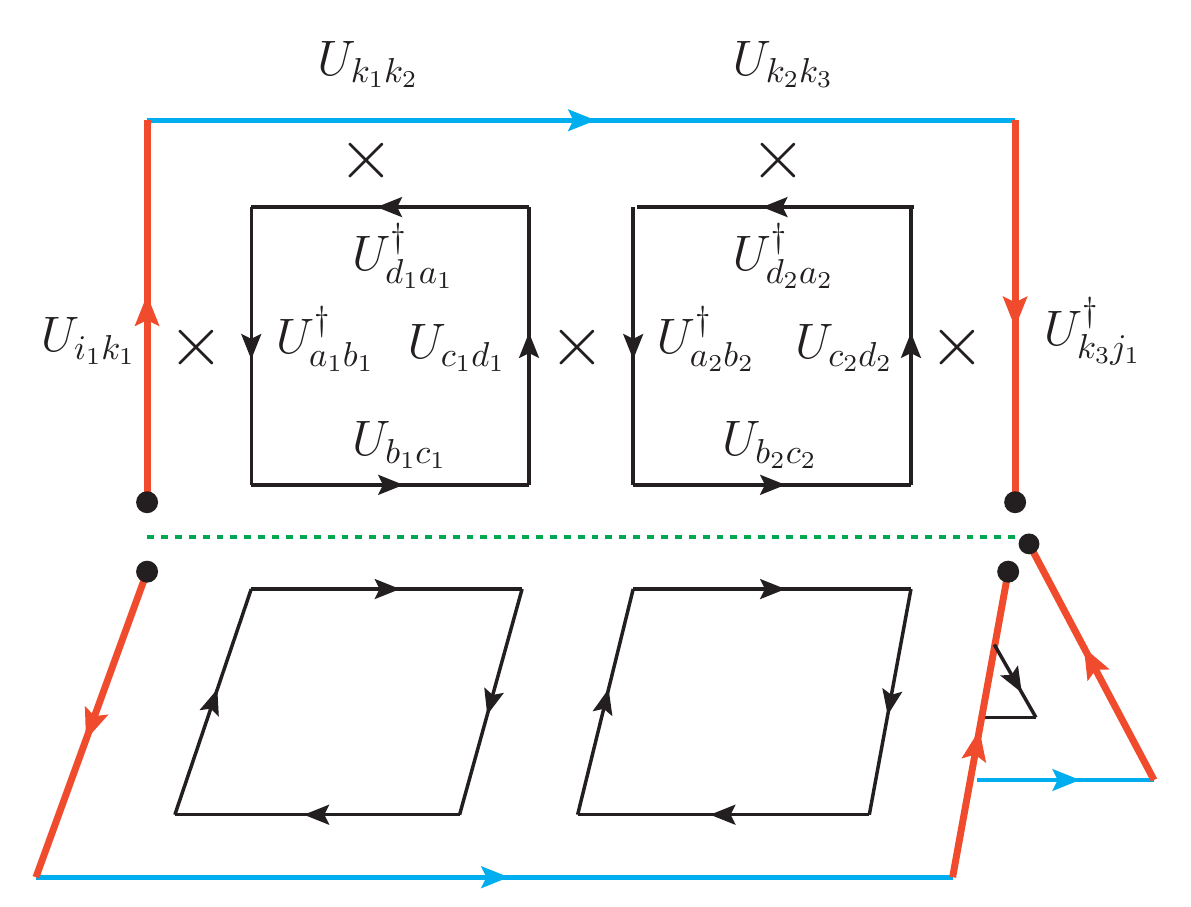}}
\caption{\small{Paving three sheets of the baryon propagator.}}
\label{fig:fig6}
\end{figure}

We want to evaluate $W_J$ in the lattice strong coupling limit~\footnote{There has been quite a number of studies of the three-quark potential in the continuum and on the lattice, starting with the seminal work of ref.~\cite{Dosch:1975gf}. Without pretending to be complete, we may mention for the study of the $\bar q\, q$ and $qqq$ potential in the continuum the work of ref.~\cite{Brambilla:1993zw} and the review~\cite{Bali:2000gf}. For the study of the $qqq$ potential on the lattice see, for instance~\cite{Alexandrou:2001ip,Takahashi:2002bw,Alexandrou:2002sn,Sakumichi:2015rfa}. }. 
As a guide for the general situation, let us consider the result of the partial calculation in which in each sheet only two plaquettes from the action are inserted (see fig.~\ref{fig:fig6}). In each sheet~\footnote{We ignore the fact that we cannot have $N$ orthogonal planes for $N>3$. This problem will be solved when rotation invariance is recovered in the continuum limit.} the five (in general $(2n_t + 1)n_s$, with $n_t$ and $n_s$ the number of plaquettes in the time and space direction, respectively, on each sheet) group integrations, marked with a cross in the figure, give the following product of Kronecker $\delta$-functions
\beq
\delta_{i_1b_1} \delta_{k_1a_1} \quad \delta_{k_1a_1} \delta_{k_2d_1} \quad \delta_{c_1b_2} \delta_{d_1a_2} \quad \delta_{d_2k_3} \delta_{a_2k_2}  \quad \delta_{d_2k_3} \delta_{c_2j_1}\, . 
\label{DELTAKR1}
\eeq
Each product of $\delta$'s that closes in a loop gives a factor $N$. In the case of the figure this means a factor $N^3$, but in general it will give a factor $N^{V-n_t-1}$ (where $V$ is the total number of vertices), and not just $V$ because the $n_t$ links along the dotted line shown in fig.~\ref{fig:fig5} (the junction) have not yet been integrated. The remaining $\delta$'s, $\delta_{i_1b_1} \delta_{c_1b_2} \delta_{c_2 j_1}$, yield the product of links $U_{i_1\ell_1}U_{\ell_1 j_1}$. This product will have to be put together with the similar products, $U_{i_2\ell_2}U_{\ell_2 j_2} \ldots U_{i_N\ell_N}U_{\ell_N j_N}$, coming from the other baryon sheets and integrated over. The result is (remember we are considering the insertion of only two plaquettes per sheet) 
\beqn
\hspace{-.8cm}&&\sum_{\ell_k}\int dU U_{i_1\ell_1} U_{i_2\ell_2} \ldots U_{i_N\ell_N} \int dU U_{j_1\ell_1} U_{j_2\ell_2} \ldots U_{j_N\ell_N} = \nn\\
\hspace{-.8cm}&&= \frac{1}{N!^2} \,\epsilon_{i_1i_2\ldots i_N} \, \sum_{\ell_k}\epsilon_{\ell_1\ell_2 \ldots \ell_N}\epsilon_{\ell_1\ell_2\ldots \ell_N} \,\epsilon_{j_1j_2\ldots j_N}= \frac{1}{N!} \,\epsilon_{i_1i_2\ldots i_N} \epsilon_{j_1j_2\ldots j_N} \, .
\label{INTERM}
\eeqn

It is clear that, in order not to get a vanishing result a new line is dynamically created along which $N$ parallel links appear. We thus see how the junction is an inevitable ingredient in the strong coupling limit of LQCD. It is also obvious that such a line can only propagate between two baryonic sources and represents the flow of the $N$ antisymmetrized colors.

To get the final result we must saturate the above tensor with the similar tensor in the definition of the baryon wave function. We thus have to put together the following factors
\begin{itemize}
\item $\frac{1}{N!} \frac{1}{\sqrt{N!}} \frac{1}{\sqrt{N!}} (N!)^2= 1$ from eq.~(\ref{INTERM}) and the normalization in eq.~(\ref{BARY})
\item $N^{-L+N_t}$ from the normalization factor of the link integration
\beq
\int dU U_{ab}U^\star_{cd}=\frac{1}{N} \delta_{ac}\delta_{bd}\, ,
\label{F1}
\eeq
\item $N^{V-N_t-1}$ from the closed loops formed around vertices with the exclusion of the vertices sitting along the junction
\item $\Big{(}\frac{1}{g^2}\Big{)}^P$ from the plaquettes  
\end{itemize}
Taking into account the presence of the $N$ sheets in the baryon expression and the Euler relation $P-L+V=2-2h-b$, we get for a surface of genus $h=0$ and $b=1$ 
\beq 
W_J=\Big{(}\frac{1}{g^2N}\Big{)}^{\sum_{k=1}^{N}P_k} \Big{[}1+\ldots\Big{]}= e^{-\kappa A_N}  \Big{[}1+\ldots\Big{]}\, ,
\label{AREALAW1}
\eeq
where $\kappa$ is the previously found string tension (see eq.~(\ref{WILDEF})) and 
\beq
A_N= |t'-t| \times \sum_{k=1}^{N} \ell_k
\label{AREALAW2}
\eeq
with $\ell_k=|\vec r_k - \vec r_J|$ the length of the curve ${\cal C}_k$ in eq.~(\ref{UDEF}). We have left unspecified the dots in eq.~(\ref{AREALAW1}). Actually, we believe that in the large-$\lambda$ limit corrections to the leading behavior are again O($1/\lambda$) or O($1/N$). In the continuum we also expect the minimal value of the quantity $\sum_{k=1}^{N} \ell_k$, hence of $A_N$, to be proportional to $N$, since, in that limit, the length of each string, $\ell_k$, will be at least of O$(\Lambda_{QCD}^{-1})$. This means that self-energy effects are expected to be of O($N$) consistently with Witten's interpretation~\cite{Witten_Baryon} of baryons as solitons in large-$N$ QCD.

We thus see that the baryonic Wilson loop has in the large-$\lambda$ limit {\it exactly} the same expression as the usual Wilson loop (see eqs.~(\ref{WGLUEB}) and~(\ref{WGLUEBA})) in terms of the 't Hooft coupling and the total area of the pages of our $N$-page ``book''. This is in agreement with the fact that each Wilson line (flux tube) is assumed to be in the fundamental representation of $SU(N)$. Furthermore, the baryon states of highest angular momentum for a given mass (the so-called leading Regge trajectory) correspond to keeping the area of two (for $N=3$) pages fixed and small and to increasing the size of the third. This is how we understand the universality of the Regge slope for $q\,\bar{q}$ mesons and baryons.

One final remark is in order. Suppose that we fix the spatial position of our three (or $N$) quarks but not that of the junction. In the strong (or large 't Hooft) coupling limit the position of the junction will be dynamically determined by the condition of minimizing the sum of the areas of the pages of the book. Let us show that this corresponds precisely to the condition that the junction is in equilibrium as a result of the $N$ forces exerted on it by the strings coming from each quark. 

In order to minimize the area we need to minimize the sum of the distances between the junction and each quark 
\beq
D \equiv \sum_1^N \sqrt{(\vec x_i - \vec x_J)^2} \, .
\eeq
Setting to zero its derivatives with respect to $x_J^k, k = 1,2, \ldots, N$, we simply get
\beq
 \sum_1^N \frac{\vec x_i^k - \vec x_J^k}{\sqrt{(\vec x_i - \vec x_J)^2}} = 0 \, ,
\eeq
which is the stated equilibrium condition since the strengths of the $N$ forces are independent of their lengths for a linear potential~\footnote{A similar argument can be found in ref.~\cite{Kalashnikova:1996px}.}. Furthermore, as argued in~\cite{Witten_Baryon}, in the heavy quark limit the junction is not moving in space. 

\subsection{Large-$N$ considerations}

In a classic paper~\cite{Witten_Baryon} Witten argued that baryons should be regarded as solitons in the large-$N$ limit of QCD.  His claim was based on the rather convincing argument that the baryon mass spectrum should be proportional to $N \sim 1/g^2$, $1/g^2$ being a typical soliton mass. Witten also pointed out that getting this result from the $N$-dependence of baryonic correlators in perturbation theory is somewhat subtle, since such a correlator actually contains arbitrarily high powers of $N$.
We have just seen an example of this in the strong coupling expansion of the baryonic Wilson loop which goes like the exponential of a sum of $N$ areas.
The correct interpretation is that the Wilson loop is related to a finite (Euclidean) time propagator $\exp(-E\tau)$ with $E \sim N$. This is, basically, the same as Witten's argument in large-$N$ perturbation theory. For the interested reader we briefly sketch (our understanding of) Witten's argument in Appendix A. 

The interesting outcome of that analysis, relevant for the present investigation, is that the large-$N$ limit of QCD {\it can} also be used for the baryonic (and more generally for the multiquark) sector.
The fact that the mass of such states goes to infinity with $N$ does {\it not} imply that large-$N$ results cannot be used at $N=3$. Let us give an example: suppose that one can prove that, at large $N$, $m_{B} = m_{M}N/2$ for a certain baryon $B$ and meson $M$. We would then argue that, to the extent that $N=3$ can be considered to be sufficiently large, it is possible to predict a ratio $3:2$ between, say, the proton and the $\rho$ meson.
More generally, certain baryonic quantities may diverge with $N$ in a precise way, so that, once such a dependence is factored out and properly taken into account, the rest is just a function of $\Lambda_{{\rm QCD}}$. Since, by definition of the large-$N$ limit, $\Lambda_{{\rm QCD}}$ is $N$ independent, one would obtain interesting estimates of those quantities at $N=3$, possibly in terms of other (e.g.\ mesonic) quantities.

\subsection {Baryon-(anti)baryon scattering: tetraquarks as baryonia}

\begin{figure}[htbp]
\centerline{\includegraphics[scale=0.5,angle=0]{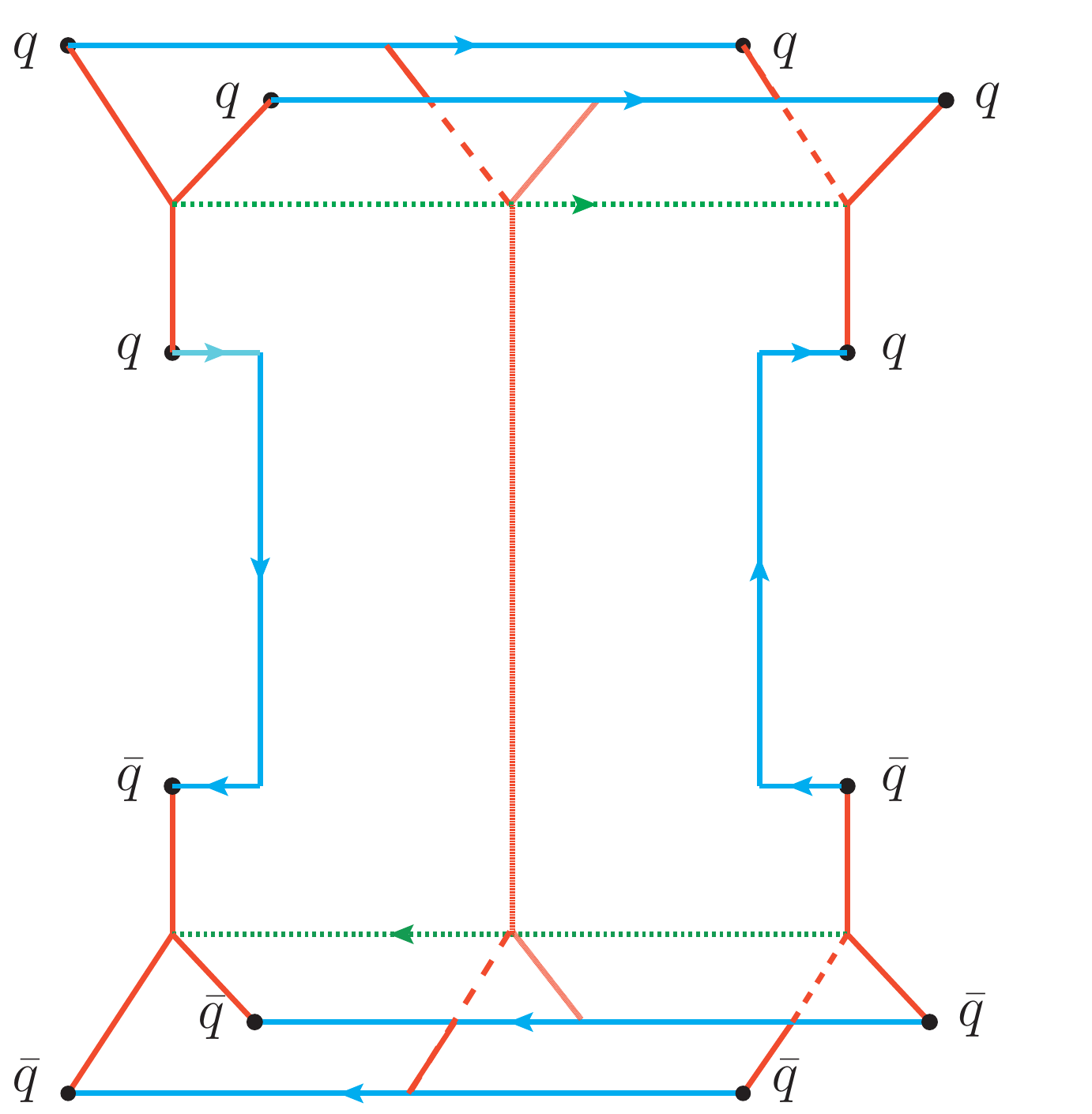}}
\caption{\small{Baryon-antibaryon scattering at large $\lambda$, showing an $s$ channel $M_4^J$ baryonium (tetraquark) state dual to a $t$ channel $q\, \bar q$ meson.}}
\label{fig:fig7}
\end{figure}

\begin{figure}[htbp]
\centerline{\includegraphics[scale=0.5,angle=0]{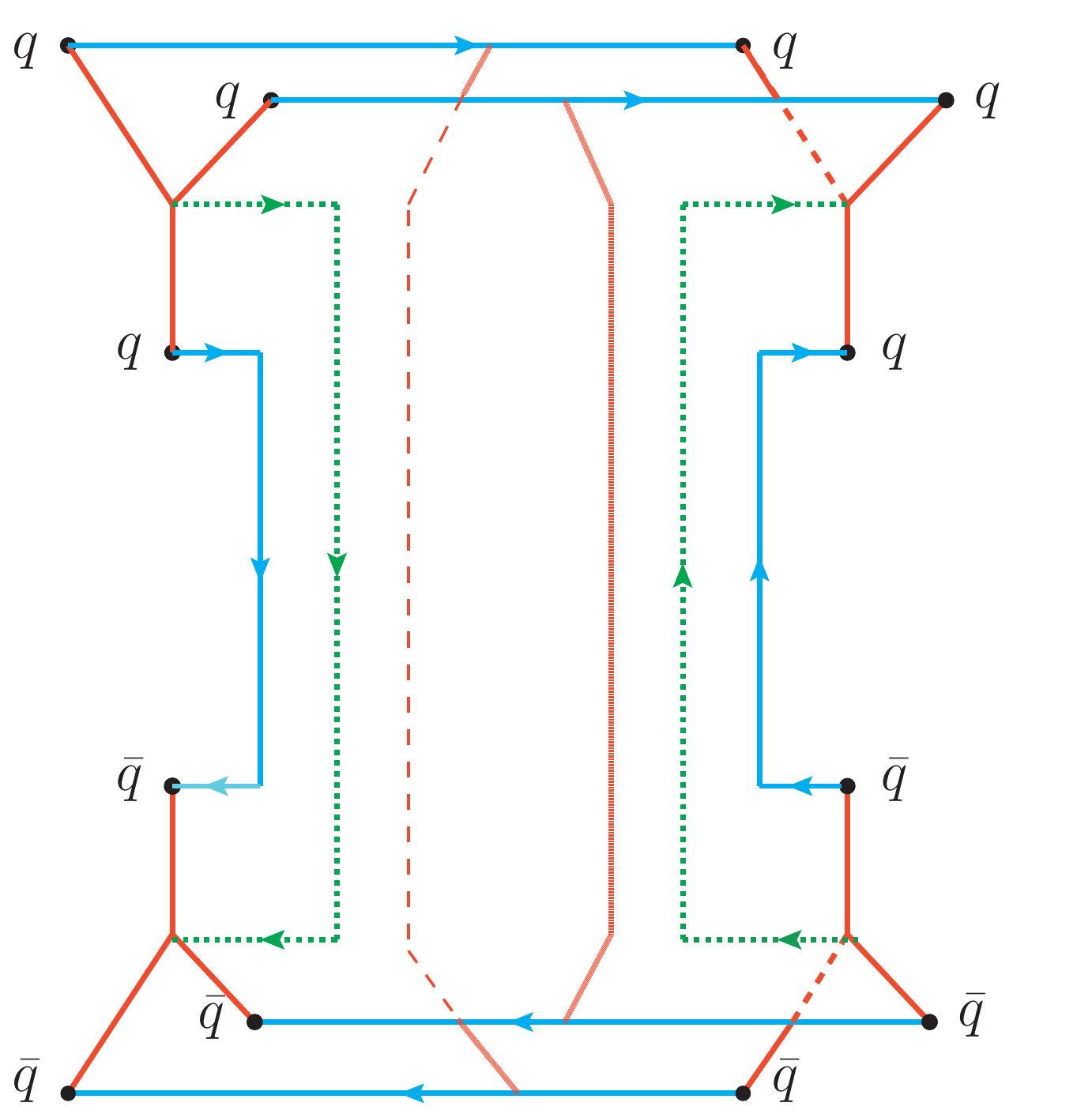}}
\caption{\small{Baryon-antibaryon annihilation at large $\lambda$, showing a pair of two $s$ channel $q\, \bar q$ mesons dual to a $t$ channel $M_2^J$ baryonium state.}}
\label{fig:fig8}
\end{figure}

It is relatively straightforward to extend the considerations of the previous subsection to baryon-(anti)baryon scattering. Since, as already mentioned, each junction necessarily connects two baryonic sources (more precisely: either an incoming and outgoing baryon or antibaryon, or an incoming or outgoing baryon-antibaryon pair) we will have {\it two} distinct strong coupling diagrams for a given flow of the quark lines. For each one of them the strong coupling limit selects a minimal-area surface with the correct topology for the quark lines and the junction. These are shown for $N=3$ in figs.~\ref{fig:fig7} and~\ref{fig:fig8} for a particular choice of the flavor (quark line) flow.

The flow of the junction lines determines which channel has also purely mesonic intermediate states (i.e.\ annihilation) and which does not (a third channel has total baryon number two and no intermediate states in this approximation). This classification of strong coupling contributions to baryon-(anti)baryon scattering makes precise, within QCD, Rosner's original observation of the necessity of tetraquarks as the intermediate states in the $s$ channel of fig.~\ref{fig:fig7} which are characterized by the fact of containing a junction-antijunction pair. These states are necessary in order to reproduce the imaginary part of the Regge amplitude corresponding to ordinary $q\,\bar{q}$ mesons in the $t$ channel~\footnote{Exchange degeneracy of the corresponding positive and negative signature trajectories is forced by the absence of a $u$(baryon number two)-channel  discontinuity.}. 

By contrast the intermediate states in the $s$ channel of fig.~\ref{fig:fig8} include two $q\,\bar{q}$ meson annihilation states. In the language of DHS duality, tetraquarks are dual to $q\,\bar{q}$ mesons, while baryon-antibaryon annihilation into two mesons is dual to a new kind of state: a $q\,\bar{q}$ state with two junctions (denoted by $M_2^J$ in~\cite{Rossi:1977cy}). To the extent that such states lie on lower Regge trajectories, the annihilation channel is suppressed (at least at high energy) with respect to the scattering channel that proceeds via tetraquark (baryonium) intermediate states. 

Naturally, more that one $q \, \bar q$ pair can annihilate, if flavor allows it. We do not repeat here the detailed discussion of all these other cases as it can be found in refs.~\cite{Rossi:1977cy,Montanet:1980te}, together with the complete classification of the large $s$, fixed $t$ behavior of all scattering and annihilation amplitudes at $N=3$. The latter is summarized in Tables~3a and~3b of ref.~\cite{Montanet:1980te}. 

We refer the reader to the specialized literature~\cite{Prelovsek:2008rf,Prelovsek:2010ty,Wagner:2012ay,Ikeda:2013vwa} for  investigations aimed at proving the existence of tetraquarks, as diquark-antidiquark states or as a molecule made of two $q\,\bar q$ singlets, in lattice simulations. 

\subsection{States with more than 2 junctions}

\begin{figure}[htbp]
\centerline{\includegraphics[scale=0.5,angle=0]{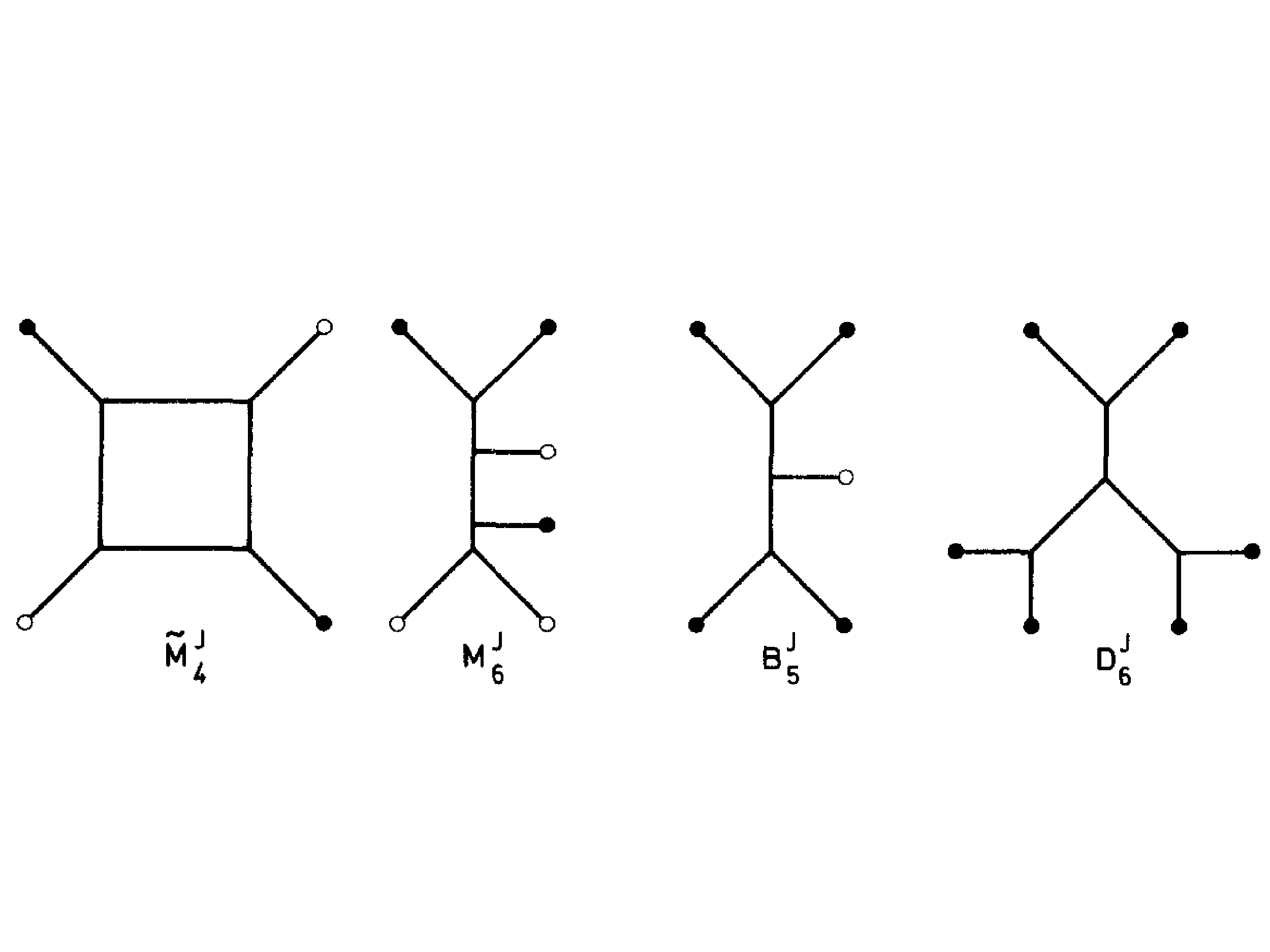}}
\vspace{-2.cm}
\caption{\small{States of the baryonium family for $N=3$ with more than two junctions.}}
\label{fig:baryonia}
\end{figure}

Arguments very similar to those of the previous section can be used to argue that also states with more than four quarks and more than two junctions should exist. The simplest ones are pentaquarks with 2 junctions and one antijunction and baryon number one. As we shall see in the next section, the JOZI rule implies that these states are mostly coupled to  $BB\bar{B}$ channels and that, when such channels are not  kinematically open, they will decay into a baryon plus mesons with small widths. Other states with more than two junctions can be constructed. Some of them are depicted, for $N=3$, in fig.~\ref{fig:baryonia} (corresponding to fig.~10 of ref.~\cite{Montanet:1980te}). 

\vspace{-.4cm}

\section {The JOZI rule}
\label{sec:JOZI} 

\subsection{Need for a selection rule on top of OZI's}

Let us start by recalling that the usual OZI rule has two important consequences for $q\,\bar{q}$ mesons:

\begin{itemize}
\item it suppresses decays which need an initial $q\,\bar{q}$ pair to annihilate and a new pair to be created; 
\item it suppresses flavor mixing in the mass matrix.
\end{itemize}
The second property is responsible, for instance, for the ``ideal mixing" of the vector meson nonet. Hence the $\phi$ meson is (almost) a pure $s \bar{s}$ state. Then, by the first property, it prefers to decay into $K \bar{K}$ rather than into pions, an experimental fact.
Of course this decay is allowed because $K$-mesons are light enough, which, in turn, is the case because of the pseudo Nambu-Goldstone (PNG) boson nature of the pseudoscalar nonet.
There is a further twist, though, to this story: the pseudoscalar sector appears to be anomalous. Its ``onia" are {\it not} ideally mixed because the  $U(1)_A$-anomaly gives large OZI-violating contribution (compared to the small mass difference of PNG bosons). Hence we do not see any sign of the OZI rule in the quark composition and decay of the $\eta'$.

Obviously, as we move to $q\,\bar{q}$ mesons containing heavier quarks ($c$ and $b$, essentially) the OZI rule comes back in all the channels. Furthermore, since charmed or beauty mesons are by no means light PNG bosons, the preferred decay of heavy quarkonia is often not allowed kinematically and therefore the lightest ones, ($J/\Psi, \Upsilon, \dots$) are narrow.
The usual OZI rule applies also, of course, to tetraquarks. However, we claim that, by itself, it cannot explain the phenomenology of tetraquark states.
Take for instance a candidate tetraquark state containing a $c\bar{c}$ pair (besides, say, up and down quarks). The usual OZI rule would suppress decays in which there is no $c$ quark in the final state, but would certainly allow decay into two charmed mesons if this is allowed by phase space.
And indeed, if we look at fig.~\ref{fig:fig9}, we see that, for instance, the $X(3872)$ putative tetraquark decays into mesons containing the charmed pair. But then, if such a decay were unsuppressed, how can we explain a width of less than 2~MeV? It looks that another selection rule should be there {\it on top} of OZI. This could be the JOZI rule that we introduced long ago and that we shall now discuss, although other explanations have been proposed~\cite{VO,RGG,TORN1,TORN2,Yang:2011wz,Karliner:2015ina}.

In analogy with the ordinary OZI rule the new rule recognizes the existence of a hidden baryon number inside tetraquarks with a $J-\bar{J}$ pair. Thus, like with hidden charm or beauty, the claim is that decay into a pair of baryons is strongly favored whenever kinematically accessible. Similarly, a pentaquark containing two junctions and an antijunction will preferentially decay into two baryons and an antibaryon.
A certain number of tetra- and pentaquark states, however, may lie below threshold for their JOZI-allowed decays. In that case, the state is expected to be narrow even if it has plenty of phase space for decaying into mesons or a baryon plus mesons, respectively. Again this is similar to the heavy quarkonium situation. Note that the JOZI rule also implies that production of narrow tetra and pentaquark states from JOZI-violating processes is suppressed. This may explain why some earlier attempts to find narrow baryonium states through direct baryon-antibaryon annihilation failed to give convincing results. Forming such states from the weak decay of heavy quarks appears to be a much better strategy.

We will now discuss, successively, some theoretical and phenomenological arguments in support of the JOZI rule.

\subsection {Supporting theoretical arguments}

Theoretical arguments supporting the JOZI rule follow closely those used in Section~\ref{sec:OZIR} to argue that the strong coupling (or large-$\lambda$) expansion justifies the ordinary OZI rule.
We should indeed compare, for a given flow of the quark lines, diagrams with or without junction annihilation like those in Figs.~\ref{fig:fig7} and~\ref{fig:fig8}. Let us first look into this question from the large-$N$ expansion viewpoint.

Consider, in general, a duality diagram in which $n$ quark lines annihilate in the $s$ channel while the remaining $(N-n)$ go through and then distinguish the case in which also the junctions go through or annihilate. One can tile the analog of the diagrams of  figs.~\ref{fig:fig7}  and~\ref{fig:fig8} with a minimal-genus set of plaquettes  and, using the usual integration rules, evaluate the large-$N$ behavior for the two diagrams. A straightforward counting gives:
\beqn
&&A^{{\rm scattering}}_{(n, N-n)} \sim N^{-n}\label{Ncounting1}\, ,\\
&&A^{{\rm annihilation}}_{(n, N-n)} \sim N^{-(N-n)}\label{Ncounting2}\, ,
\eeqn
which is consistent with the fact that the same diagram (rotated by 90 degrees) describes both scattering and annihilation provided we also change $n$ into $(N-n)$.

The above equation shows consistency with the usual OZI rule for the scattering case (each annihilation costs an extra factor $1/N$). However, for annihilation, the equation seems to favor a large number of annihilating $q\, \bar{q}$ pairs. Also, comparison between the two processes (for the same $n$) shows that scattering dominates at $n < N/2$ while annihilation dominates at $n > N/2$. However, before jumping to premature conclusions, we should note that extra combinatorial factors related to the number of possible ways in which a given diagram can be built, may have to be taken into account. Such a number involves also the flavor structure of the process and is, in general, quite complicated.

\begin{figure}[htbp]
\centerline{\includegraphics[scale=0.7,angle=0]{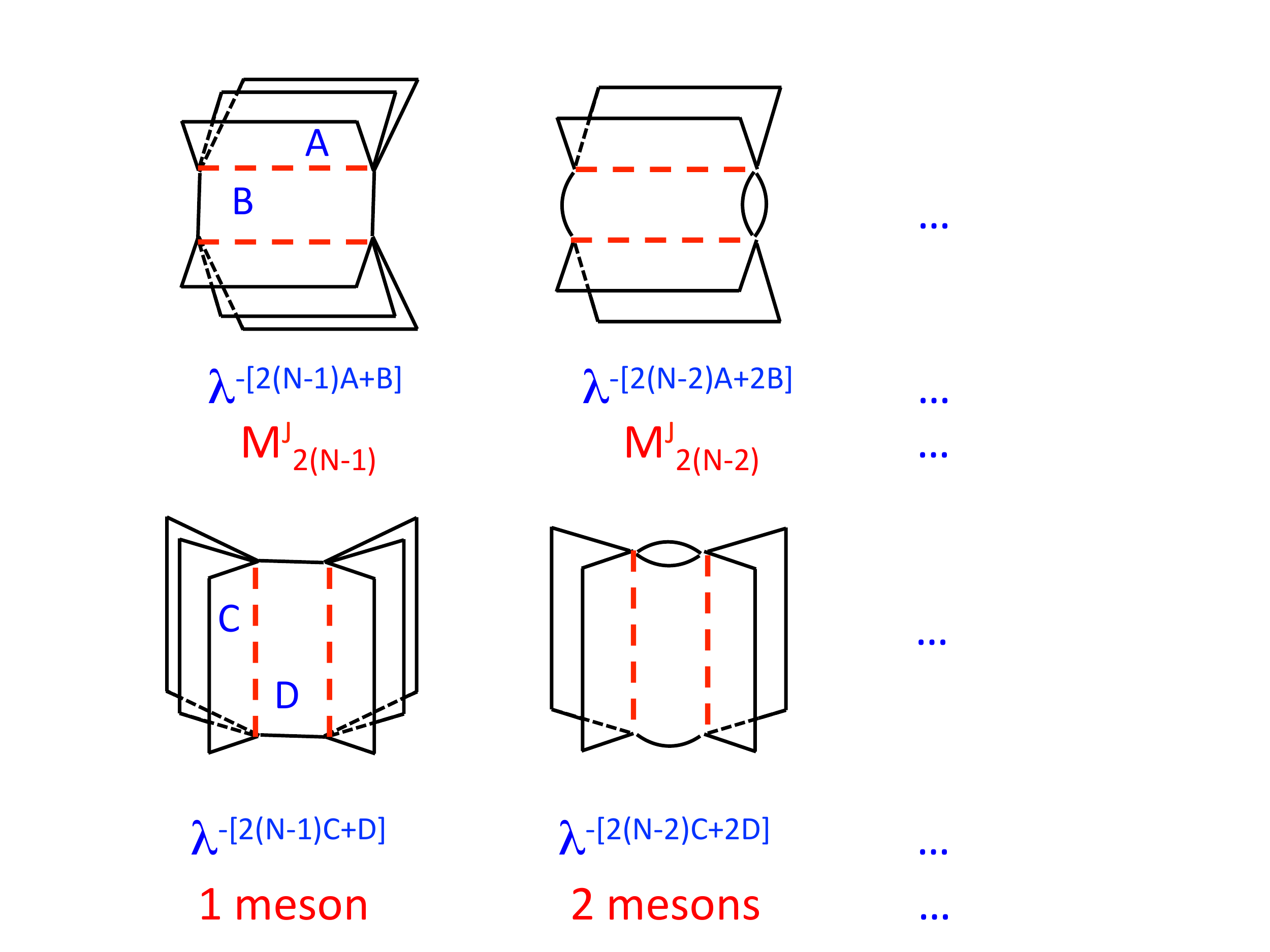}}
\caption{\small{The $s$ channel intermediate states of scattering (top panels) and annihilation diagrams (bottom panel) and the corresponding $\lambda$ behavior.}}
\label{fig:fignew}
\end{figure}

If we take a large-$\lambda$ limit we have to compare minimal area surfaces for the two processes. The following large-$\lambda$ behaviors appear
 \beqn
&&A^{{\rm scattering}}_{(n, N-n)} \sim \lambda^{- 2 N A + n (2 A -B)} \, ,\label{lambdacounting1}\\
&&A^{{\rm annihilation}}_{(n, N-n)}  \sim \lambda^{- 2 N C + (N-n) (2 C -D)} \, ,\label{lambdacounting2}
\eeqn
where the areas $A,B,C,D$ are those indicated in fig.~\ref{fig:fignew}. In that figure we schematically indicate the $s$ channel intermediate states of scattering (top panel) and annihilation diagrams (bottom panel) as well as the leading $\lambda$ behavior of each of them.

Again, a large $n$ implies many extra sheets in the scattering diagram while a large $(N-n)$ calls for many sheets in the annihilation diagram (i.e.\ for for the $(N-n)$ mesons). This fits well with the large-$N$ result but falls short of justifying the JOZI rule in general. 
At this point we can combine the above arguments with the one already given in~\cite{Rossi:1977cy} where it was pointed out that, at high energy, the diagram in fig.~\ref{fig:fig7} dominates over the one in fig.~\ref{fig:fig8} owing to the exchange of a higher Regge trajectory (ordinary meson vs.\ a baryonium state). Such an argument can be repeated in the general case discussed above and supports both the ordinary and the junction OZI rule independently of $n$. We see here an analogy with what we found already in Section~\ref{sec:JFS} for the validity of the ordinary OZI rule. Sometimes the large-$N$ and high-energy arguments go hand in hand whereas in other cases they appear to push in opposite directions. 

Another theoretical support for the JOZI rule can be based on a judicious extrapolation from $N=3$ to large $N$. Here our approach can be contrasted to Weinberg's recent one~\cite{SWEIN}. His large-$N$ considerations are based on keeping the quark-gluon content of the tetraquark unchanged as $N$ is increased. In our framework, instead,  one should decide first on how to extrapolate the $N=3$ tetraquark state to arbitrary $N$. Obviously, the state should contain two junctions but then how many quarks? One possible extrapolation would consist in keeping the number of quarks fixed at four and simply increase to $(N-2)$ the number of strings connecting the two junctions. At the opposite extreme one might prefer to keep a single string stretched between the junctions and have $2(N-1)$ quarks. Or, finally, one may like to increase both the number of quarks and the number of intermediate strings keeping their ratio fixed while $N \rightarrow \infty$. In all cases the mass of such a state is expected to grow linearly with $N$.

One can then estimate for each case the $N$-dependence of the decay amplitudes into different channels with and without junctions and compute the $N$ dependence of the relative branching ratios. Without going into the details of a straightforward though tedious calculation, one finds that the dominant transitions (of O$(N^{-1/2})$) are cascades in which a single quark-antiquark pair is created or annihilated while preserving the junction structure. Furthermore, the decay amplitude into a baryon-antibaryon pair goes like $N^{-n/2}$ if there are $n$ intermediate strings that have to be broken. Annihilation amplitudes decrease for increasing $N-n$ (and thus decreasing $n$), but are always exponentially suppressed in agreement with~\cite{Witten_Baryon}. Therefore, quite generically, the ratio between the JOZI-violating and JOZI-conserving partial widths goes quickly to zero at large $N$. Assuming $N=3$ to be large enough, the tetraquark state should also be mesophobic (and narrow) in real QCD.

In conclusion, in spite of several arguments in favor, a general justification of the JOZI selection rule from either large-$N$, large-$\lambda$, or high-energy considerations is still lacking. In the end, one has to appeal to the data in order to check its validity or usefulness. This is indeed the point we wish to address now.

\subsection{Supporting phenomenological arguments}

The overall pattern of masses and widths of candidate multiquark states is phenomenologically quite puzzling as, generally speaking, they look anomalously long-lived. This, by the way, is precisely the reason that makes them visible as resonances over the strong-interaction continuum. 

In order to discuss the situation, we found convenient to collect in figs.~\ref{fig:fig9} and~\ref{fig:fig10}, respectively, the masses and widths of today's available (and more or less confirmed) hidden charm and hidden bottom mesonic (putative tetraquark) states. A similar compilation for baryonic (putative pentaquark) states is presented in fig.~\ref{fig:fig11}. 

\subsubsection{Mesonic states: candidate tetraquarks}

\begin{figure}[htbp]
\centerline{\includegraphics[scale=0.6,angle=0]{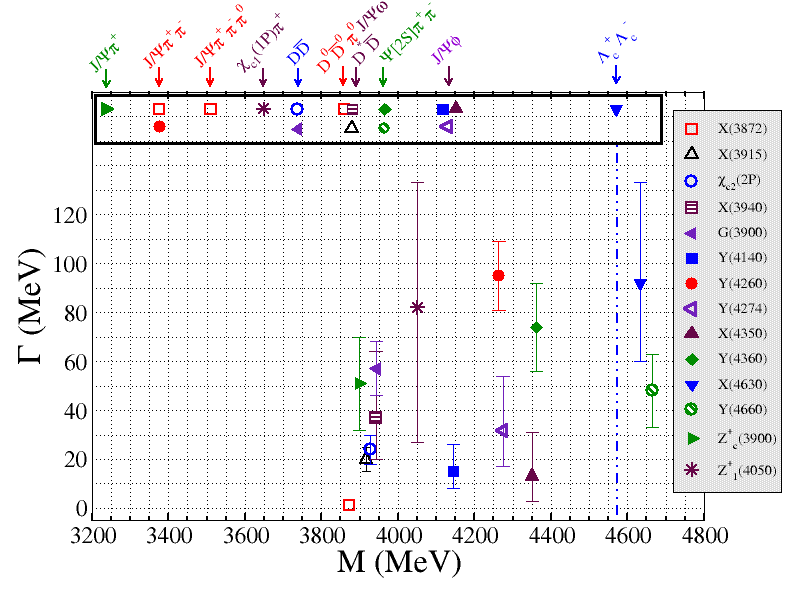}}
\caption{\small{Masses and widths of narrow hidden charm candidate tetraquarks, taken from the compilation of ref.~\cite{Bodwin:2013nua}. We have excluded just a couple of states of that compilation whose width exceeds 100~MeV. In the upper part of the diagram we show the thresholds for the channels into which each one of the particle is known to decay. The vertical line indicates the position of the $\Lambda_c \bar \Lambda_c$ threshold.}}
\label{fig:fig9}
\end{figure}

\begin{figure}[htbp]
\centerline{\includegraphics[scale=0.6,angle=0]{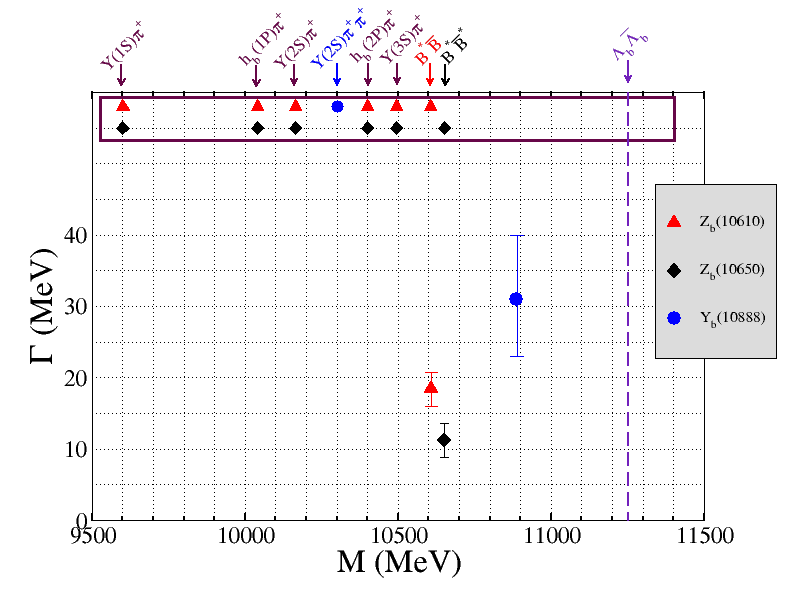}}
\caption{\small{Masses and widths of narrow candidate hidden bottom tetraquarks, taken from the compilation of ref.~\cite{Bodwin:2013nua}. We have excluded just a couple of states of that compilation  whose width exceeds 100 MeV. In the upper part of the diagrams we show the thresholds for the channels into which each one of the particle is known to decay. The vertical line indicates the position of the $\Lambda_b \bar \Lambda_b$ threshold.}}
\label{fig:fig10}
\end{figure}

The data displayed in the plots of figs.~\ref{fig:fig9} and~\ref{fig:fig10} are taken from the nice compilation of ref.~\cite{Bodwin:2013nua}. For each state, together with the mass (horizontal-axis) and width (vertical-axis),  we indicate with arrows the thresholds of the corresponding decay channels. The vertical line gives the location of the baryon-antibaryon threshold, i.e. of the $\Lambda_c \bar \Lambda_c$ and $\Lambda_b \bar \Lambda_b$ total mass, respectively.

In the case of hidden charm candidate tetraquarks (fig.~\ref{fig:fig9}) we note that, in all but one case, the allowed decay channels are purely mesonic and that there is often generous phase space allowed for the decay. The exception is the $X(4630)$ meson whose mass is a mere 60~MeV above the $\Lambda_c \bar \Lambda_c$ threshold. Yet the state is pretty large and likes to decay into charmed baryons. There is a possibility that this state is not distinct from the nearby $Y(4660)$. That case has been discussed in detail in ref.~\cite{Cotugno:2009ys} with the conclusion that the branching ratio into charmed baryons is two orders of magnitude larger than the one in $\Psi(2S) + \pi^+ \pi^-$ in spite of the huge phase space unbalance \footnote{We thank A.\ Polosa for bringing ref.~\cite{Cotugno:2009ys} to our attention.}.

Many of the states below the baryonic $\Lambda_c \bar \Lambda_c$ threshold (like $X(3872)$, $Z_c^+(3900)$, $G(3900)$, $X(3915)$, $\chi_{c2}(2P)$, $X(3940)$) are very narrow, in spite of the large available bosonic phase space. Three others ($Z^*_1(4050)$, $Y(4260)$, $Y(4360)$), though below the $\Lambda_c \bar \Lambda_c$ threshold, are somewhat larger than the six mentioned before. 

Here one should also mention the molecular interpretation of some of the states reported in fig.~\ref{fig:fig9}. In particular the states $X(3872)$, $X(3915)$, $X(3940)$ are narrow and very near to the $DD^*$ threshold as predicted by the molecular picture~\cite{VO,RGG,TORN1,TORN2,Yang:2011wz,Karliner:2015ina}. 

The situation of hidden bottom candidate tetraquarks (fig.~\ref{fig:fig10}) is even more striking. All the known states are below the $\Lambda_b \bar \Lambda_b$ threshold, but have plenty of phase space to decay into bosonic channels, yet their width to mass ratio is extremely small. 

This being said, we should make a disclaimer: we do not pretend that {\it all} narrow tetraquark states are baryonia. A more complete phenomenological study is certainly necessary before such a suggestion can be made. The above-mentioned $X(4630), Y(4660)$ tetraquark(s) do smell like baryonia. Instead, states such as $Z_b(10610), Z_b(10650)$, although narrow and below baryonic decay channels, appear to have very surprising branching ratios into open and hidden beauty channels~\cite{Bondar:2013nea} particularly after the difference in available phase space is taken into account~\footnote{We are grateful to M.\ Karliner for bringing this important point and ref.~\cite{Bondar:2013nea} to our attention.}. As we shall discuss later, the junction picture would rather predict a democratic decay into the two categories of mesonic final states. An interpretation of those states as molecules made of two heavy B-mesons appears to be possible according to present data.

\subsubsection{Baryonic states: candidate pentaquarks}

\begin{figure}[htbp]
\centerline{\includegraphics[scale=0.6,angle=0]{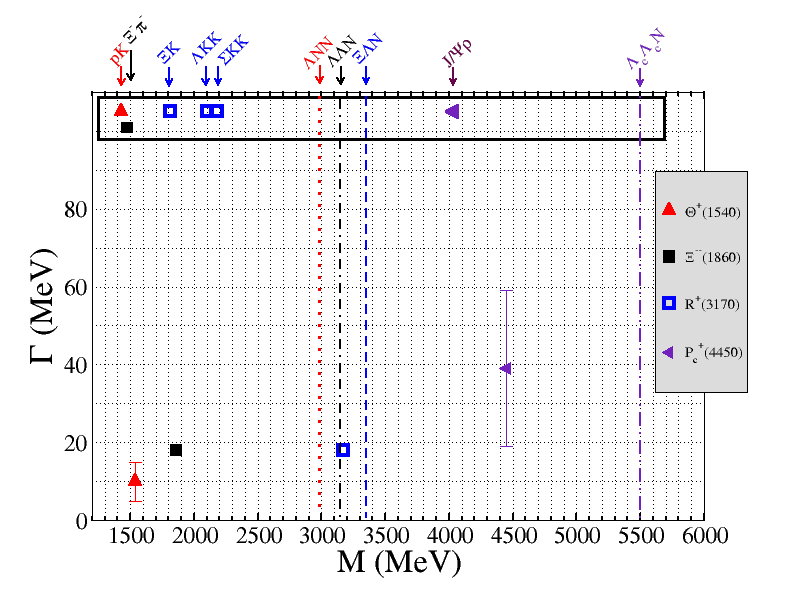}}
\caption{\small{Masses and widths of narrow candidate pentaquarks. In the upper part of the diagrams we show the thresholds for the channels into which each one of them is known to decay. The vertical lines indicate the position of the various baryon-antibaryon-baryon thresholds. Data are taken from refs.~\cite{Amirzadeh:1979qi}~\cite{Nakano:2003qx}~\cite{Alt:2003vb}~\cite{Aaij:2015tga}.}}
\label{fig:fig11}
\end{figure}

In fig.~\ref{fig:fig11} we show masses and widths of candidate pentaquark states. We put in the figure not only the narrow state recently discovered at LHCb~\cite{Aaij:2015tga}, but also the less recent (and not fully confirmed) states of refs.~\cite{Nakano:2003qx} and~\cite{Alt:2003vb}, as well as the old state discovered in 1979 again at CERN~\cite{Amirzadeh:1979qi} whose existence was actually never disproved. 

We remark that all these states have masses below the corresponding baryon-antibaryon-baryon threshold, namely $\Lambda \bar N N$ for $\Theta^+(1540)$, $\Lambda \bar\Lambda N$ for $\Xi^{--}(1860)$, $\Xi \bar\Lambda N$ for $R^+(3170)$ and $\Lambda_c \bar\Lambda_c N$ for $P_c^{+}(4450)$. Despite the fact that they have large phase space for decaying into a baryon plus bosons, they are very long-lived.

\subsubsection{Scattering amplitudes with exotic intermediate states}

If we neglect the baryon-baryonium sector of QCD (which we can at $N = \infty$),  scattering amplitudes with exotic quantum numbers should exhibit no baryonium-like resonances (they may have instead molecular bound states, but these are not our concern here). However, at $N=3$ nothing forbids exotic baryonium resonances to be formed. According to our previous discussion they are expected to be narrow whenever their JOZI-conserving decays are kinematically forbidden. We shall now illustrate this point in two examples.

\vspace{0.2cm}
\underline{Meson-meson scattering} 
\vspace{0.2cm}

Consider meson-meson scattering for the case in which the flavor content of the incoming mesons is such that $q\, \bar{q}$ annihilation is impossible or strongly  suppressed. In the absence of baryons and $J-\bar{J}$ mesons, such a scattering amplitude should have no resonances in the direct channel with its duality diagram only showing resonances in the $t$ and $u$ channels (see discussion at the end of Section~\ref{sec:JFS}). However, although with a JOZI-suppressed amplitude, a $J-\bar{J}$ pair can be created and a tetraquark can be produced. If formed, such a tetraquark will decay into a precise linear combination of two-meson states corresponding to the two ways in which the two $q\, \bar{q}$ pairs can be reconstructed.

\begin{figure}[htbp]
\centerline{\includegraphics[scale=0.55,angle=0]{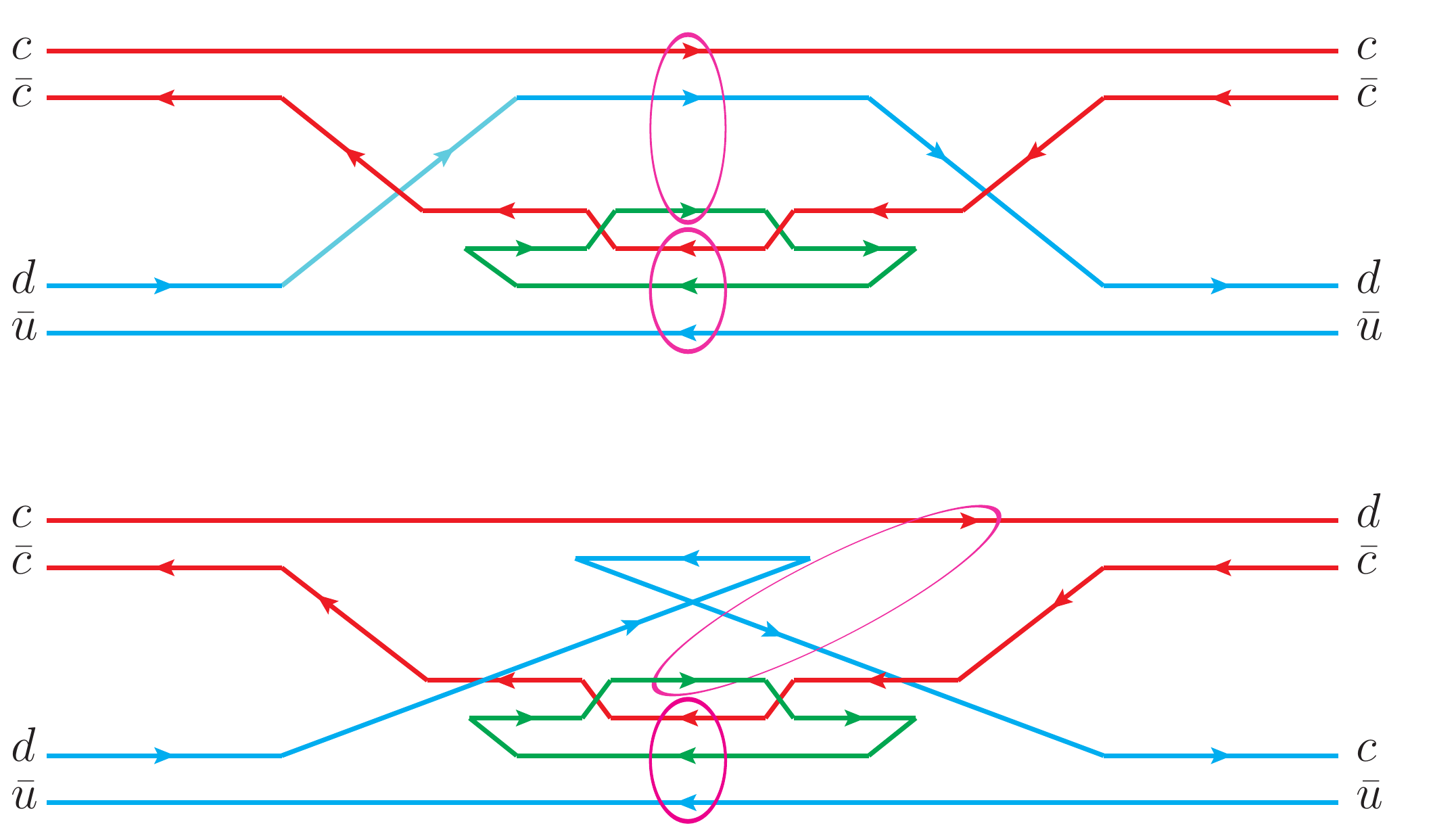}}
\caption{\small{Diagrams describing the reactions $J/\psi\, \pi^- \to J/\psi\, \pi^-$ (top panel) and $J/\psi\, \pi^- \to D^0 D^-$ (bottom panel) with an intermediate tetraquark/baryonium state. Besides the quark lines also some intermediate gluons are shown (in double-line notation) in order to evidentiate the two sets of totally antisymmetrized colors (each one encircled by an ellipse) identifying the intermediate tetraquark state.}}
\label{fig:fig12}
\end{figure}

A particularly interesting case, already considered in refs.~\cite{MPPR1,MPPR2}~\footnote{We wish to thank L.\ Maiani and A.\ Polosa for discussions on this issue.}, has a $c \,\bar{c}$ as well as a $\bar{u} \,d$ pair. Thus an intermediate tetraquark state with a $J-\bar{J}$ will be coupled, at the quark-level, to a particular combination of two states, the first consisting of a $c \,\bar{c}$  (e.g.\ $J/\psi$) and a $\bar{u} \, d$ (e.g.\ $\pi^-$) meson, the second of a $c\, \bar{u}$  (e.g.\ $D^0$) and a $\bar{c} \,d$ (e.g.\ $D^-$) meson. Two processes, with the same initial state, are represented in the top and bottom panels of fig.~\ref{fig:fig12}. The amplitudes (and thus also the rates) for the two kinds of final states  ($J/\psi\, \pi^-$ or $D^0 D^-$) should be simply related if the reaction proceeds via a single tetraquark intermediate state (each ellipse encircles three totally antisymmetrized color indices showing that a junction-antijunction pair is created and then annihilated).

\vspace{0.2cm}
\underline{Meson-baryon scattering}
\vspace{0.2cm}

\begin{figure}[htbp]
\centerline{\includegraphics[scale=0.6,angle=0]{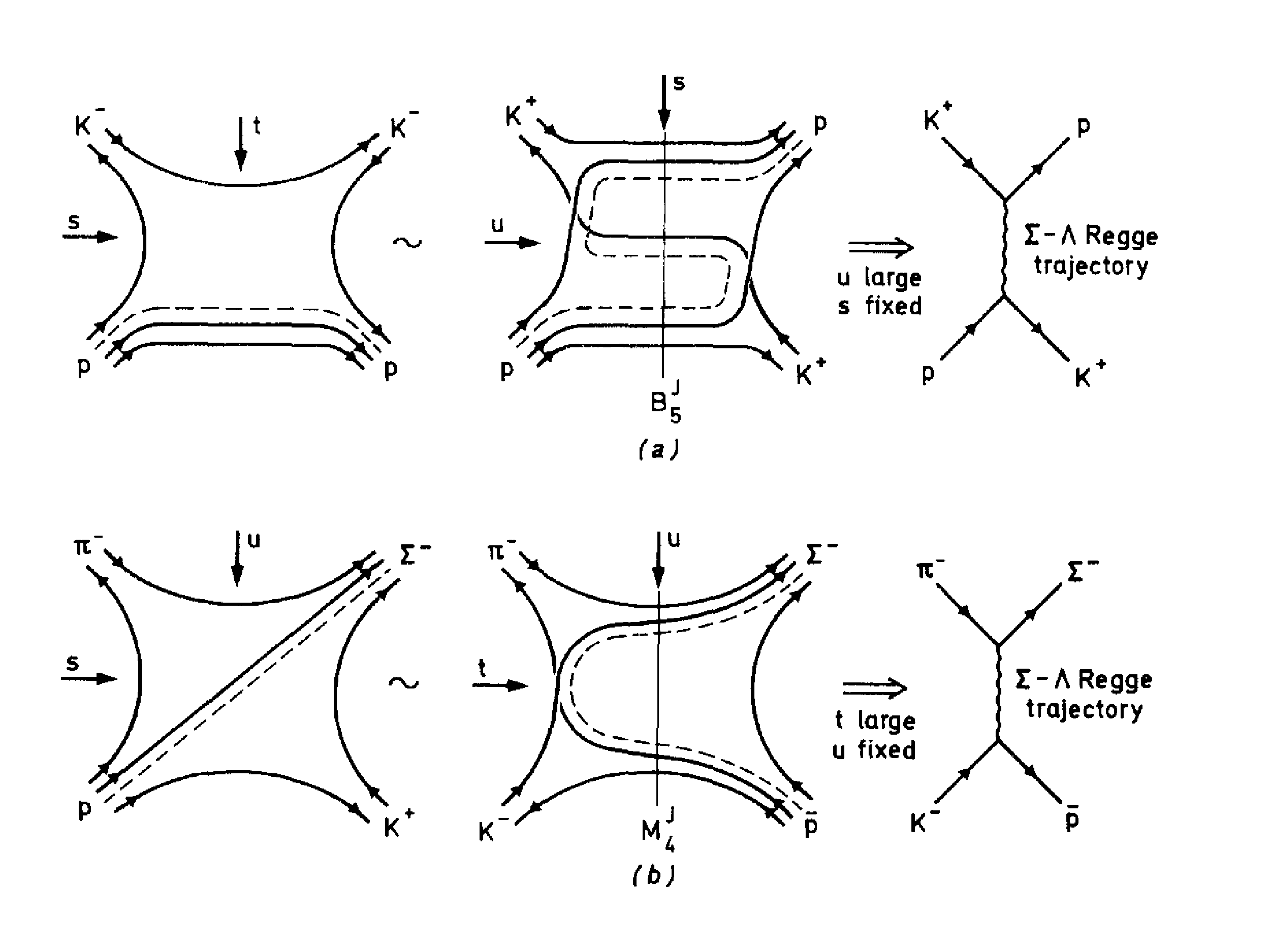}}
\caption{\small{In the upper and lower panel we provide examples of $MB\to MB$ processes where intermediate pentaquark and tetraquark states are required in order not to have a vanishing discontinuity in the $u$ and $t$ channel, respectively.}}
\label{fig:fig13}
\end{figure}

Problems with DHS duality of the kind pointed out by Rosner~\cite{ROSNER} for $B\bar B$ scattering amplitudes, also arise in the case of $MB\to MB$ processes. In the $B\bar B$ case, the exchange of baryonium states in the $s$ channel is necessary in order to account for the full $t$ channel meson trajectory (a JOZI conserving process).
Looking at fig.~\ref{fig:fig13} (taken from ref.~\cite{Montanet:1980te}), one recognizes that in the $MB\to MB$ amplitude intermediate states with junctions (pentaquarks in the upper panel and tetraquarks in the lower panel) are required if one wants to have a non-vanishing discontinuity in the $u$ and $t$ channel. respectively (JOZI violating exchange). 
Absence of these discontinuities (at least if other unitarity corrections due to more complicated, non planar, topologies are negligible) would imply exchange degeneracy for the strange baryonic Regge trajectories, $\Lambda$ and $\Sigma$, something which is phenomenologically problematic.

\section{Conclusions}
\label{sec:CONC} 

In this paper we have tried to update, both theoretically and phenomenologically, our old proposal~\cite{Rossi:1977cy} for interpreting narrow multiquark states as having hidden baryon number (``baryonia").
These states were predicted to be "mesophobic"  i.e.\ with a preference for baryonic rather than mesonic decay channels (in analogy with heavy quarkonia). This idea can be made more precise, in QCD, by introducing the concept of a junction (resp.\ antijunction) as a sink (resp.\ source) of triplets of Wilson lines, each seen as a string-like color flux tube. Single hadron states are then associated with irreducible color singlet operators containing quarks, antiquarks, junctions and antijunctions connected by Wilson lines. Besides the baryons themselves (one junction states) the simplest new states in this sector are tetraquarks and pentaquarks with two and three junctions (and baryon number zero and one), respectively. In our picture, however, other two and three-junction states with fewer quarks should also be present in the spectrum, although they might be hard to distinguish from ordinary states with the same flavor.

On the theoretical side we have discussed this picture in two distinct limits: the 't Hooft large-$N$ limit (at fixed number of flavors and fixed $\lambda \equiv g^2 N$) and a lattice strong coupling limit that we argued to be actually a large-$\lambda$ limit similar to the one often discussed in the AdS/CFT correspondence~\cite{Maldacena:1997re,Witten:1998qj}. Both limits have been successfully used for the study of the mesonic sector of QCD but, as first pointed out by Witten~\cite{Witten_Baryon}, a suitably defined large-$N$ limit can also be defined for baryons and baryonia in spite of the fact that certain quantities diverge in that limit. We have briefly reviewed Witten's approach and then argued that also the large-$\lambda$ limit can be usefully applied to the baryonic sector. 

\begin{figure}[htbp]
\centerline{\includegraphics[scale=0.6,angle=0]{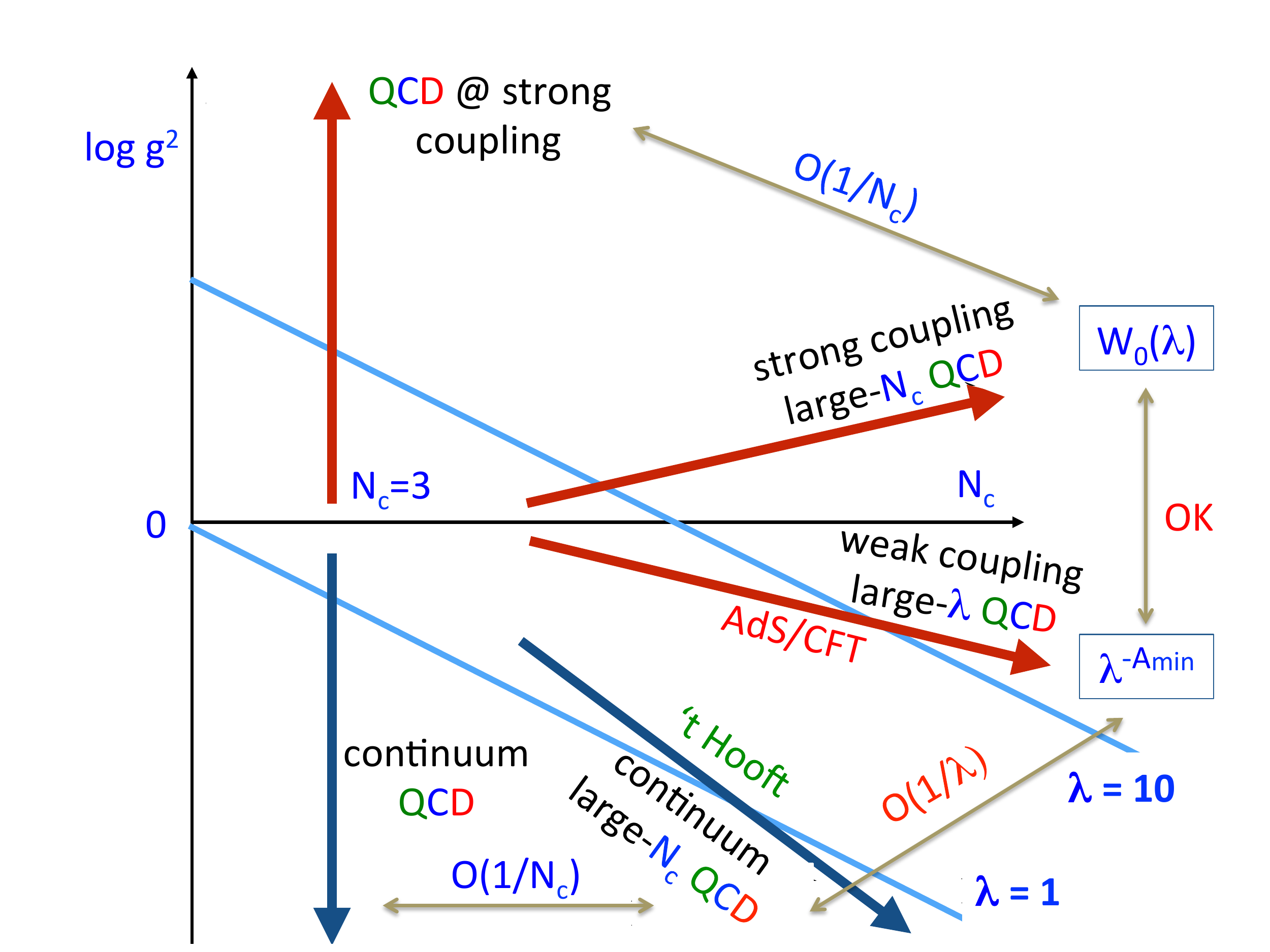}}
\caption{\small{Different regimes in (lattice) QCD. One may try to extract information about the real world ($N_c=3$ in the continuum limit) by going clockwise around a large semicircle in this diagram.}}
\label{fig:fig14}
\end{figure}

We are of course aware of the fact that both the large-$N$ and the large-$\lambda$ limits could give misleading information about real QCD: the former because $N=3$ may not be large enough, the second because the continuum limit corresponds to the $\lambda \rightarrow 0$ limit as a result of asymptotic freedom.
In fig.~\ref{fig:fig14} we show how these different limits are defined in a two-dimensional parameter space. It is somewhat reassuring that the strong coupling limit can be turned into a large-$\lambda$ one, but this is still not sufficient even to reach the continuum limit of large-$N$ QCD.

This theoretical analysis provides us with some qualitative understanding of the spectrum of multiquark states and of their interactions including the possible validity of a junction-OZI (JOZI) rule generalizing the ordinary OZI rule to the baryonic/baryonium sectors. In spite of these considerations the validity of the JOZI rule appears to depend on a number of extrapolations that may be satisfied for some states and not for others.

We think that only a detailed study of the relevant experimental data can tell whether the string-junction picture for multiquark states is useful. In this paper we have undertaken a first and modest step in this direction by collecting present data about multiquark states, their masses, widths and decay channels.
The gross features that strikes the eye is that many of these states are unusually narrow even when they have considerable phase space for decaying into channels that violate the JOZI. All of them are below threshold for decaying into JOZI-allowed channels with one exception, the $X(4630)$ state, which is just above the $\Lambda \bar{\Lambda}$ threshold and yet is pretty broad.

We believe that, all in all, our forty years old proposal appears to hold water, but a more detailed study of the existing experimental situation data, as well as the acquisition of new data, is necessary before discriminating our approach from several others that are presently on the market. It is quite likely that there is no single explanation for the existence of narrow multiquark states. We have argued that the fifty year old DHS duality, as well as several QCD approximations (large-$N$, large-$\lambda$), predict the existence of tightly bound ``irreducible" multiquark states. But, of course, QCD should better not exclude that in proton-neutron scattering a narrow ``molecule", usually called deuteron, will appear. It is perfectly possible that, similarly, some tetra and pentaquark states are narrow, not because of the OZI or JOZI rules, but simply because they are ``molecules" whose mass is close to the  dissociation threshold into their constituent ``atoms".

\vspace{.5cm}

\section*{Acknowledgements} 
We wish to thank P.\ Dimopoulos for helping us with some of the figures and G.\ Ricciardi for bringing some recent experimental data to our attention. One of us (GV) has enjoyed  a number of interesting discussions about the strong coupling and large-$N$ limits with J.-B.\ Zuber as well as enlightening correspondence on the molecular picture with M.\ Karliner. 
We also acknowledge useful exchanges with M.\ Bochicchio, L.\ Maiani, A.\ Polosa, E.\ Querchigh and C.\ Sonnenschein. 

\appendix 
\renewcommand{\thesection}{A} 

\section{Sketch of Witten's large-$N$ argument}

In principle, to study the large-$N$ limit of baryon dynamics one should set up a relativistic and gauge-invariant $N$-particle bound state equation, a highly not trivial task. In practice, being only interested in some qualitative conclusion, one can employ the following reasoning: one starts by separating the $N$ quark interaction potential $V^{(N)}$ from the free $N$ quark propagator. This is done, technically, by associating the former with $N$-quark-irreducible (NQI) diagrams (diagrams that do not have an intermediate state consisting of just $N$ dressed quark propagators). Next, one  decomposes $V^{(N)}$  into contributions from two-body, three-body, etc.\ interactions all the way up to an $N$-body term (a set of fully connected diagrams). In formulae we write
\beq
V^{(N)} = \sum_{n = 2}^N V_n^{(N)}\, .
\label{V(N)}
\eeq
Finally, one uses the large-$N$ limit to analyze each term in the above sum. Eventually, the energy levels of the system will be given by the eigenvalues $V^{(N)}$.

\begin{figure}[htbp]
\centerline{\includegraphics[scale=0.8,angle=0]{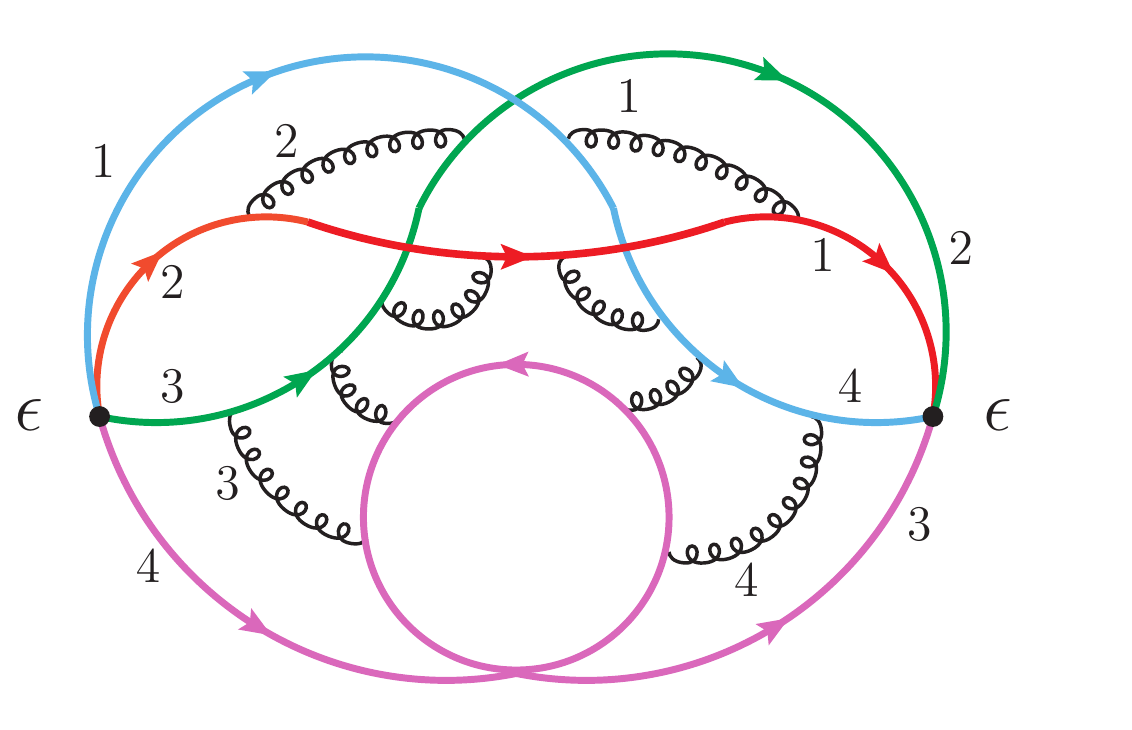}}
\caption{\small{Typical leading (``planar'') diagram contributing to the baryon wave function for $n=N=4$. Unlike in the rest of the paper different colors represent different flavors. The numbers indicate how the 4 colors flow in the diagram.}}
\label{fig:fig15}
\end{figure}

By its definition $V_n^{(N)}$ consists of diagrams that have a connected (amputated) $2n$-point function accompanied by $N-n$ non-interacting (but dressed) propagators. Our claim is that, at leading order in $1/N$, the above mentioned connected $2n$-point function can be constructed out of an $n$-meson planar amplitude by the following procedure. Take all the planar diagrams giving the leading contribution to the correlation function of $n$ gauge-invariant quark-antiquark bilinears. Then, open up the quark bilinears, and, instead, simply join the $n$ quark and $n$ antiquark lines with the  remaining $(N-n)$ ``spectator" quarks and antiquarks at two vertices inserting at each of them an overall $N$-dimensional Levi-Civita $\epsilon$ factor. This is illustrated in fig.~\ref{fig:fig15} for  $n=N=4$, where one can easily visualize the above mentioned  $4$-meson amplitude by joining together the $q-\bar{q}$ pairs sharing the same color label (i.e. the same number since the different colors actually denote the quark's flavor).

In order to determine the $N$-dependence of $V_n^{(N)}$ one will have to sum over all possible inequivalent ways of applying the above procedure. This will produce an ${N \choose n}$ binomial coefficient (and not its square since the $n$ quark colors have to match those of the $n$ antiquarks) as well as a factor $(n-1)!$ from the number of inequivalent cyclic + anticyclic orderings of the pairs in the $n$-meson diagram. Finally, there will be a factor $g^{2(n-1)} \sim N^{-(n-1)}$ from the $n$-meson amplitude itself. Summing up we find
\beqn
 &&V_n^{(N)} \sim N^n N^{-(n-1)} F(n,\lambda) = N F(n, \lambda)\, ,  \label{V(N)1}\\
 &&V^{(N)} =  N \sum_{n = 2}^N  F(n,\lambda)  \, , \label{V(N)2}
\eeqn 
where $F(n,\lambda)$ includes  some of the above-mentioned combinatorial factors as well as the dynamical $n$-dependence of the $n$-boson correlator. Note, however, that the $F(n,\lambda)$ coefficients {\it do not} depend upon $N$. As a result, provided their $n$ dependence is not too singular (i.e. that the sum over $n$ in (\ref{V(N)1}) converges fast enough), $V^{(N)}$ will be simply  proportional to $N$ with an overall universal coefficient. The above considerations make somewhat more precise the conditions under which Witten's ``soliton claim" follows. It also confirms his point that  the problem of determining the relevant quark interaction inside a baryon at large-$N$ is of the same level of difficulty as that of finding the large-$N$ mesonic scattering amplitudes. 

\end{document}